\documentclass[12pt]{article}

\usepackage{times}
\usepackage{graphicx}           
\usepackage{epstopdf}
\usepackage{subfigure}
\usepackage{color}
\usepackage{caption}
\usepackage{cite}

\title{\vspace{-1cm}Attosecond spectroscopy of size-resolved water clusters}
\author
{X. Gong$^{1,2,\ast}$, S. Heck$^{1,\ast}$, D. Jelovina$^{1}$, C. Perry$^{1}$, K. Zinchenko$^{1}$, H. J. W\"{o}rner$^{1,\dagger}$ \\
\normalsize{$^1$ Laboratorium f\"{u}r Physikalische Chemie, ETH Z\"{u}rich, 8093 Z\"{u}rich, Switzerland}\\
\normalsize{$^{2}$ State Key Laboratory of Precision Spectroscopy, East China Normal University, Shanghai, China}\\
\normalsize{$\ast$ These authors contributed equally to this work.}\\
\normalsize{$\dagger$ e-mail: hwoerner@ethz.ch}
}
\date{}
\begin{document} 

\baselineskip24pt

\maketitle 

\begin{abstract}
 Electron dynamics in water are of fundamental importance for a broad range of phenomena\cite{sanche09a,boudaiffa00a,garrett05a}, but their real-time study faces numerous conceptual and methodological challenges\cite{svoboda20a,loh20a,jordan20a}. Here, we introduce attosecond size-resolved cluster spectroscopy and build up a molecular-level understanding of the attosecond electron dynamics in water. We measure the effect that the addition of single water molecules has on the photoionization time delays\cite{cavalieri07a,schultze10a,kluender11a,ossiander16a,huppert16a} of water clusters. We find a continuous increase of the delay for clusters containing up to 4-5 molecules and little change towards larger clusters. We show that these delays are proportional to the spatial extension of the created electron hole, which first increases with cluster size and then partially localizes through the onset of structural disorder that is characteristic of large clusters and bulk liquid water. These results establish a previously unknown sensitivity of photoionization delays to electron-hole delocalization and reveal a direct link between electronic structure and attosecond photoemission dynamics. Our results offer novel perspectives for studying electron/hole delocalization and its attosecond dynamics.

\end{abstract}


Electronic dynamics in water play a central role in a broad range of scientific and technological research areas ranging from radiation chemistry to photocatalysis. The dynamics induced by ionization of water are of particular relevance since they initiate the processes underlying radiation damage\cite{boudaiffa00a,garrett05a,alizadeh15a}. The ionization of water is predicted to lead to the formation of a delocalized electron hole, followed by its localization on one water molecule and proton transfer to a neighboring molecule, forming H$_3$O$^+$ and OH\cite{marsalek11a}. The latter step has been time-resolved only very recently using one-photon extreme-ultraviolet (XUV) photoionization of water clusters\cite{svoboda20a} and strong-field ionization of liquid water\cite{loh20a}. Both experiments independently determined a 30-50 fs time scale for proton transfer. The formation of the delocalized electron hole, as well as its localization have so far escaped experimental scrutiny because of their sub-femtosecond time scales.

In this work, we access the attosecond time scale of the photoionization dynamics of water on the molecular level by introducing attosecond size-resolved cluster spectroscopy (ASCS). Coupling attosecond interferometry\cite{kluender11a,huppert16a,cattaneo18a,nandi20a} with electron-ion coincidence spectroscopy, we determine photoionization delays for water clusters of increasing size, achieving single-molecule resolution. Photoionization time delays of (H$_2$O)$_n$ are found to continuously increase from $n=1$ to $n=4-5$. We show that this increase directly reflects the augmenting delocalization of the electron hole created in the ionization process. For these small clusters, we find a linear relationship between the photoionization time delays and the first moment of the electron-hole density created in the ionization process. Beyond $n=4-5$ the photoionization delays vary little, an effect that we attribute to the partial localization of the electron hole caused by the onset of structural disorder characteristic of larger clusters and bulk liquid water. These assignments are further confirmed by calculations on the O-1s photoionization delays of water clusters, which display these effects even more clearly owing to the atomic character of the orbitals. As we show below, the present results also confirm the interpretation of photoemission delays from liquid water \cite{jordan20a}.

Our work thus also reveals a possible experimental access to the spatial delocalization of electronic wave functions, which has always been difficult to characterize. Electron delocalization plays a fundamental role in the properties of solids, where the perfect translational symmetry of single crystals creates fully delocalized electronic wave functions (or Bloch waves), which are disrupted by local disorder in a phenomenon known as Anderson localization\cite{anderson1958,anderson_bk97}. The delocalization of electronic wave functions is also central for understanding the aromaticity of molecules, charge transfer between a metal atom and its ligands, or between a solute and a solvent. The electronic structure of water clusters and liquid water has so far mainly been accessible through quantum-chemical calculations, which have predicted partial (de)localization of the electronic wavefunctions\cite{prendergast05a}. Experimental access to this information has however not been reported so far.

\begin{figure}
    \centering
    \includegraphics[width=1.0\textwidth]{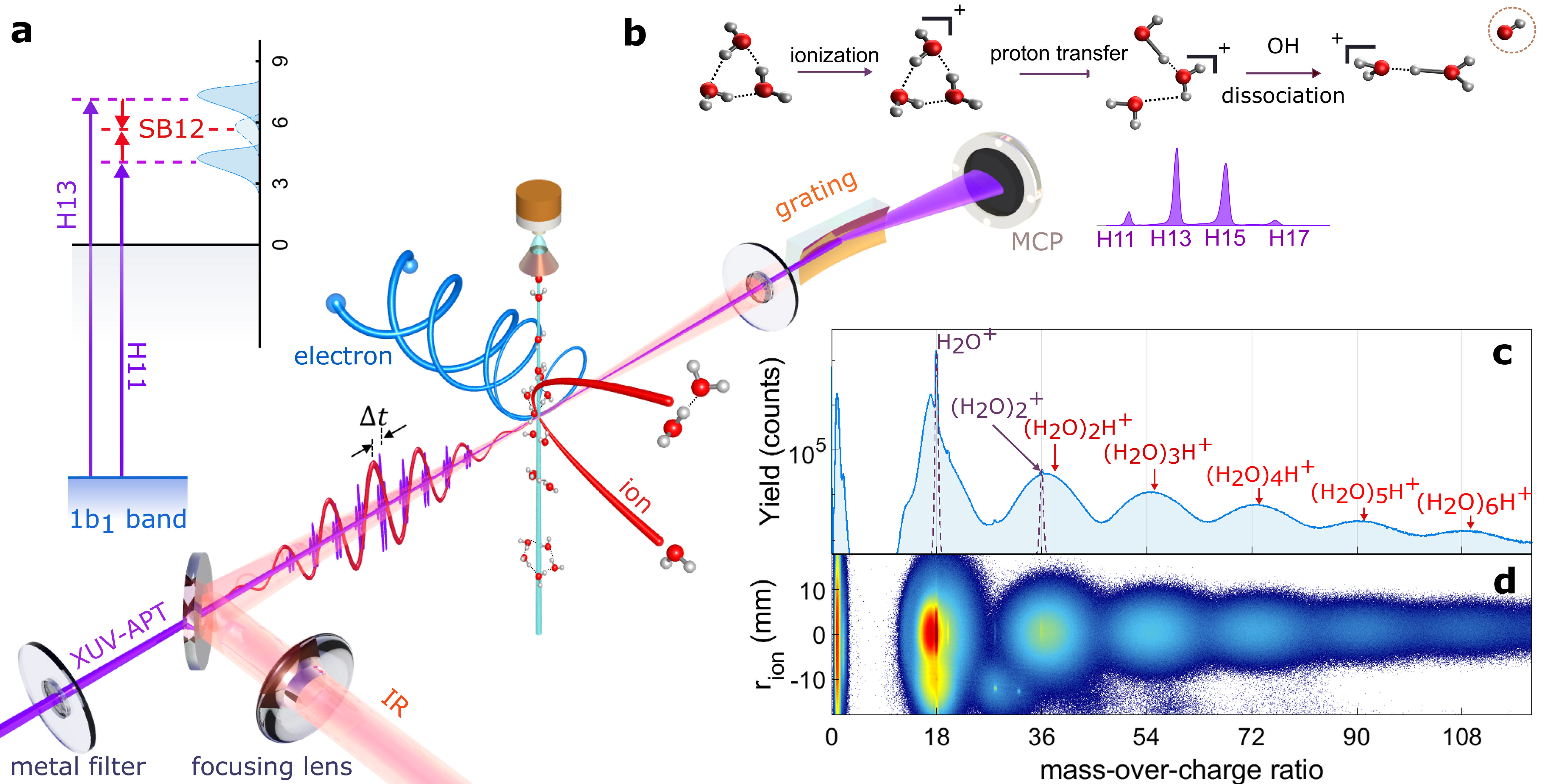}
    \caption{
    \textbf{Attosecond size-resolved cluster spectroscopy.} \textbf{a}, Experimental setup: The charged cluster fragments and emitted photoelectrons are measured in coincidence as a function of the APT-IR delay. The XUV APT is characterized via an online soft-X-ray spectrometer. The inset on the left illustrates ionization by two neighboring harmonic orders and the creation of the sideband spectrum. \textbf{b}, Dissociative-ionization mechanism of the water trimer, illustrating the only relevant fragmentation pathway for all observed cluster sizes. \textbf{c}, Mass spectrum of the cluster beam photoionized by an APT as a function of the mass-over-charge (MOC) ratio. \textbf{d}, Two-dimensional MOC spectrum of water cluster species as a function of position on the ion detector.
    }
    \label{setupcluster}
\end{figure}

Figure 1 provides a conceptual overview of our measurements. An XUV attosecond pulse train (APT) generated through high-harmonic generation is focused into a supersonic water-cluster beam, where it is spatio-temporally overlapped with a near-infrared (IR) laser pulse. The APT and IR pulses are phase locked through an actively-stabilized Mach-Zehnder interferometer. The three-dimensional momentum distributions of electrons and ions generated from this interaction are detected in coincidence using COLd Target Recoil Ion Momentum Spectroscopy (COLTRIMS)\cite{Dorner2000a, Jagutzki2002,Ullrich_2003} [see Methods for details]. The photoionization time delays of water clusters are measured by recording photoelectron spectra as a function of the time delay between the overlapping APT and IR pulses, in coincidence with each ionic fragment. As shown in the inset of Fig.~1a, single-photon XUV ionization gives rise to the main bands (MB) in the photoelectron spectra, whereas the additional IR interaction creates sidebands (SB).

The unique assignment of the coincident attosecond photoelectron spectra to a specific cluster size is possible because of a dissociative-ionization mechanism that is general for small ($<$ $\sim$20 molecules) water clusters at low ionization energies (Fig. 1b). Following outer-valence single ionization, water clusters undergo rapid proton transfer, followed by the loss of a single OH unit on a sub-picosecond time scale\cite{Dong2006,SHIROMARU1987,Shi1993,Wei1994,Bobbert2002,Tachikawa2004,Belau2007,Liu2011,Zamith2012}, such that each detected fragment (H$_2$O)$_n$H$^+$ mainly originates from the neutral (H$_2$O)$_{n+1}$ precursor for n $<$ 6 (see SM Section 1.2 for details).
The observed mass spectrum (Fig. 1c) indeed shows a well-resolved progression of broad peaks that is easily assigned to (H$_2$O)$_n$H$^+$ with $n=2-6$. The width of the peaks is caused by the kinetic-energy release in the dissociative photoionization, as highlighted in Fig. 1d, which shows the mass spectrum as a function of the detected position radius of the ions on the detector ($r_{\rm ion}$).
The only unprotonated species (H$_2$O$^+$ and (H$_2$O)$_2^+$, purple dashed curves) originate from the photoionization of H$_2$O and (H$_2$O)$_2$, respectively. The broad distribution peaking at a MOC of 17 is OH$^+$ originating from the dissociative ionization of H$_2$O$^+$. A fraction of the photoionized dimers remains bound, leading to the sharp (H$_2$O)$_2^+$ peak, and the remainder dissociates to produce (H$_2$O)H$^+$. Analogous results have been obtained following the ionization of D$_2$O clusters. The corresponding mass spectra are shown in Fig. S7.

\begin{figure}
    \centering
    \includegraphics[width=1.0\textwidth]{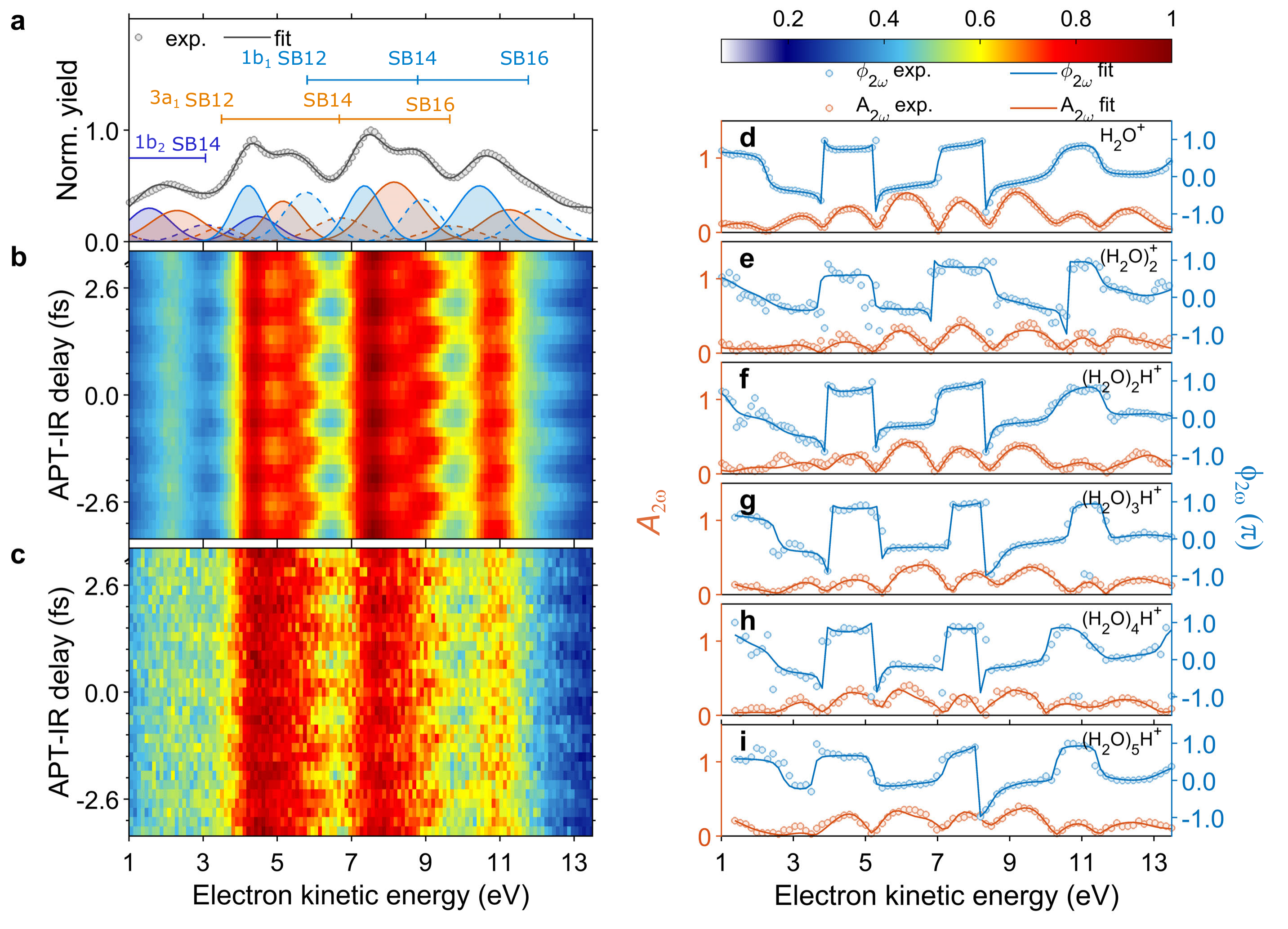}
    \caption{
    \textbf{Attosecond photoelectron spectroscopy of size-selected water clusters.} Attosecond photoelectron spectra created by overlapping XUV-APT and IR pulses and detected in coincidence with H$_2$O$^+$, integrated over the APT-IR delay \textbf{a} and shown as a function of APT-IR delay \textbf{b}. \textbf{c}, Same as \textbf{b}, but detected in coincidence with (H$_2$O)$_2$H$^+$. The false-color map for \textbf{b} and \textbf{c} is shown in the top right corner.  \textbf{d} to \textbf{i}, The Fourier transforms at 2$\omega$ of the attosecond photoelectron spectra detected in coincidence with  \textbf{d} H$_2$O$^+$, \textbf{e} (H$_2$O)$_2^+$,  \textbf{f} (H$_2$O)$_2$H$^+$, \textbf{g} (H$_2$O)$_3$H$^+$, \textbf{h} (H$_2$O)$_4$H$^+$, \textbf{i} (H$_2$O)$_5$H$^+$, shown in terms of their modulation amplitude ($A_{2\omega}$, orange color) and phase ($\phi_{2\omega}$, blue color). The experimental data and the fitted curves are shown as open circles and solid lines, respectively.
    }
\end{figure}

Figure 2 shows the attosecond photoelectron spectra (APS) obtained in coincidence with each cluster size. The APS measured in coincidence with H$_2$O$^+$ (Fig. 2a) is dominated by the contributions of harmonic orders 11, 13, and 15, and ionization from the two outermost (1b$_1$ and 3a$_1$) molecular orbitals of H$_2$O. The black line shows a fit using the literature values of the vertical binding energies. The filled spectra correspond to the decomposition of the APS in MB (full colored lines) and SB (dashed lines) spectra. Figures 2b and 2c show the characteristic oscillations with a period of 1.33~fs in the APS coincident with H$_2$O$^+$ and (H$_2$O)$_2$H$^+$, respectively. The remaining APS are shown in Fig. S3. Analogous results obtained for D$_2$O clusters are shown in Fig. S8. 
The SB-intensity oscillations take the form $A_{q} \propto A_{2\omega,q}\cos(2\omega(\tau - \tau_{q}))+B_q$, where $\omega$ is the angular IR frequency, $\tau$ is the experimentally varied APT-IR delay, $\tau_{q}=\tau_{q}^{\rm XUV}+\tau_{q}^{\rm sys}$, $\tau_{q}^{\rm XUV}$ is the harmonic emission time, and $\tau_{q}^{\rm sys}$ is the system-specific photoemission time delay. Here, we determine relative photoemission delays between water clusters (H$_2$O)$_n$ and H$_2$O, as a function of $n$, which cancels the contribution of $\tau_{q}^{\rm XUV}$. Because the ionization energies vary by less than 0.6~eV from $n=1-6$, the relative measurement also causes negligible contributions of the continuum-continuum (or Coulomb-laser coupling) delays\cite{dahlstrom12a,pazourek15a} on the order of 4-6~as for SB12-14.

The main challenge in the determination of photoionization time delays from such measurements is the considerable spectral overlap. We therefore use a general procedure, introduced\cite{jordan18b} and validated in our recent work\cite{jordan20a},  that resolves this challenge. Instead of integrating the APS oscillations over specific spectral regions, our approach fully accounts for the spectral overlap. Briefly, we Fourier transform the APS along the time-delay axis and then fit the complex-valued Fourier transform at the 2$\omega$ frequency by assigning a specific phase shift to each spectral component of the MB and SB spectra. Details are given in the SM, Section 1.5. We keep the spectral positions and amplitudes fixed to values determined from the delay-integrated spectra (see, e.g., Fig. 2a and S3). The success of this fitting procedure is highlighted by the excellent agreement between the experimental data (circles in Figs. 2d-i) and the fits (full lines). The robustness of the fitting procedure to variations of the initial guesses is shown in Fig.~S6.

\begin{figure}
    \centering
    \includegraphics[width=1.0\textwidth]{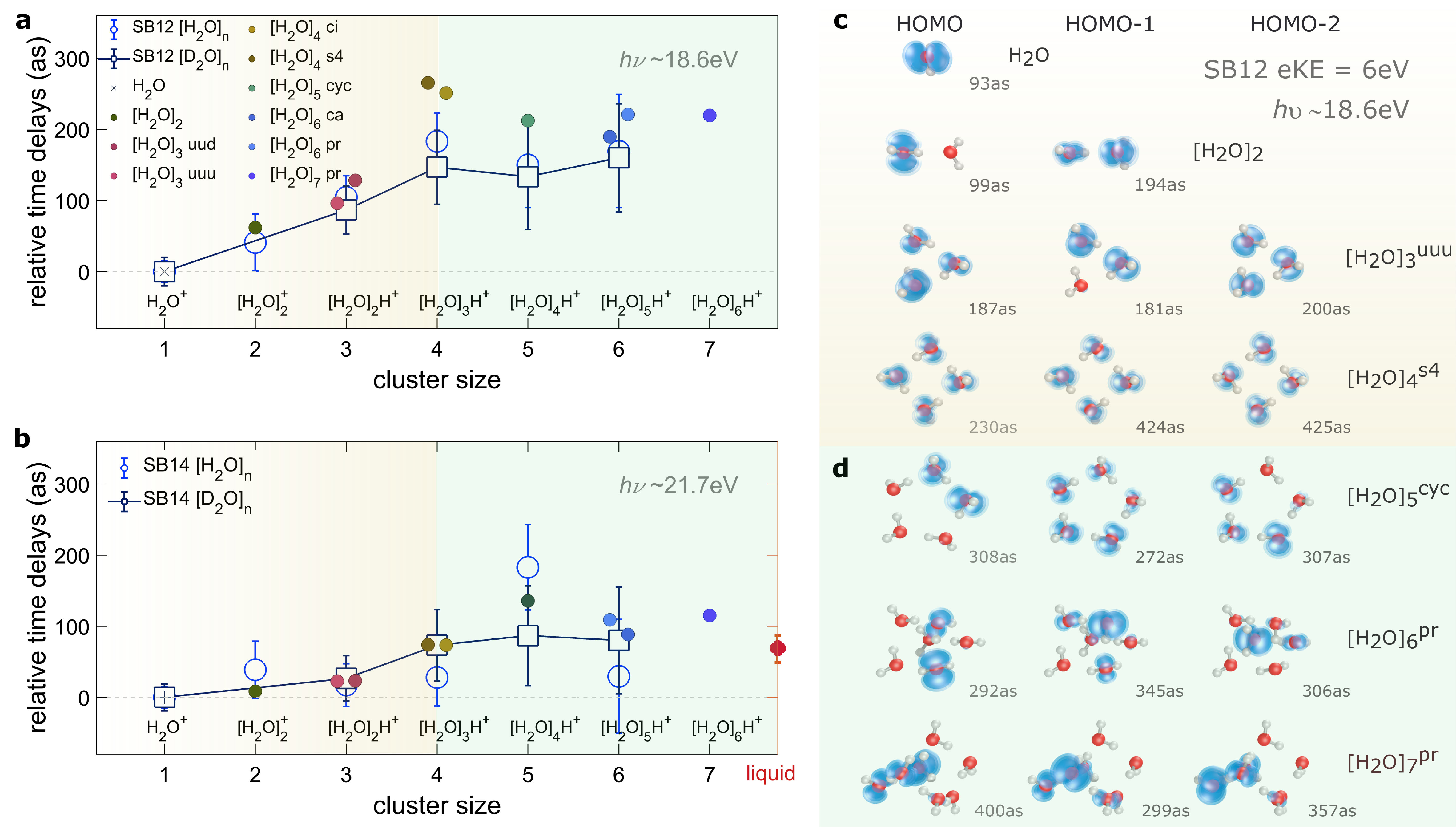}
    \caption{
    \textbf{Size-resolved photoionization time delays of water clusters.} \textbf{a}, Time delays for photoionization out of the 1b$_1$ band of water clusters, relative to H$_2$O (or D$_2$O), measured in  SB12 (empty circles). The error represents the combined uncertainty from the timing jitter and standard deviation. The calculated delays (filled symbols) were obtained for a kinetic energy of 6.0 eV. \textbf{b}, Same as a, but measured for SB14, or calculated with eKE = 9.1~eV, additionally showing the relative photoemission time delay of liquid water reported in Ref.\cite{jordan20a}. \textbf{c}, Electron density map and calculated absolute photoionization delays of the 1-3 highest-occupied orbitals of the 1b$_1$ band of water clusters, highlighting the effect of delocalization (orange shading), followed by partial localization (\textbf{d}, green shading).
    }
\end{figure}

Figures 3a and 3b show the cluster-size-resolved photoionization time delays corresponding to the 1b$_1$ photoelectron bands from monomer to hexamer, relative to the monomer delay, as determined from SB12 and SB14, respectively. The time delays measured in SB12 (18.6~eV photon energy) increase as a function of the cluster size up to the tetramer, followed by little variation. The results for SB14 show a similar behavior, with indications of a slightly slower convergence as a function of cluster size. The latter results can be compared to our recent measurements of bulk liquid water, which yielded a photoemission delay of 69$\pm$20~as relative to the water monomer for SB14\cite{jordan20a}, indicated as the red dot in Fig. 3b. The close agreement between the H$_2$O- and D$_2$O-cluster results suggest that nuclear-motion effect are not relevant within the accuracy of the present measurements.

To understand the mechanisms governing these delays, we performed {\it ab-initio} quantum-scattering calculations of the photoionization delays (see SM Section 2 and Ref.\cite{baykusheva17a} for details). Starting from the trimer, each water cluster exists in several isomeric forms\cite{shi93h2on,Temelso2011a,Malloum2019,Liu1996,Liu1996a,Xantheas2000,Richardson2016,cvitas20a}. At the low temperatures reached in our supersonic expansion, only one or two isomers are thermally populated, as detailed in the SM (Section 1.2 and Table S1). We used the equilibrium geometries of the most stable cluster isomers reported in Ref.\cite{Temelso2011a} to perform electronic-structure calculations with a correlation-consistent valence-triple-zeta (cc-pVTZ) basis set. These served as an input to the photoionization calculations performed by solving the electron/water-cluster-ion scattering problem at the experimentally relevant scattering energies using the iterative Schwinger variational principle\cite{gianturco94a,natalense99a}. The input orbitals, scattering potentials and scattering wave functions were all represented by single-center expansions using a typical maximal angular-momentum value of $\ell=50$, whereby numerical convergence with respect to this parameter was ensured. The photoionization time delays, resolved as a function of photoemission direction in the molecular (cluster) frame and the cluster orientation in the laboratory frame, were obtained and subsequently angularly averaged using the partial photoionization cross sections as weighting factors (for details, see SM Section 2). These calculations yielded angular-integrated one-photon-ionization (or Wigner) delays. We have compared these delays to two-photon (XUV+IR or RABBIT) delays obtained by additionally including the effect of the IR field on the photoionization delays. The results, shown in Figs. S9-S13, establish the close correspondence of the one- and two-photon delays in the case of water clusters, both angle-resolved and angle-integrated, and therefore support our comparison of the angle-averaged one-photon delays with the observables of the present experiment. 
Using this methodology, photoionization delays were obtained for each of the $n$ orbitals of (H$_2$O)$_n$ that contribute to the 1b$_1$ band of each water cluster.
The cross-section-weighted average of these delays (defined in Eq.(1) in the Methods section) are shown as the large filled symbols. The agreement between theory and experiment is excellent (Figs. 3a and 3b).

Figures 3c and 3d show the densities of the highest-occupied molecular orbitals of the 1b$_1$ band of the most stable isomer of each cluster size, together with their absolute photoionization delays. Figure 3c suggests that the increasing orbital delocalization correlates with the increasing time delay. The HOMO of the dimer has almost the same time delay as the monomer, whereas the delocalization of the HOMO-1 in the dimer leads to an increase of the time delay by nearly 100 as. This comparison also shows the very small effect ($\sim$ 6~as) of electron scattering on the neutral H$_2$O neighbor in the dimer. Among the trimer orbitals, it is also the most delocalized orbital (HOMO-2) which has the largest photoionization delay (200~as). The tetramer orbitals are perfectly delocalized over all molecules, owing to the S$_4$ symmetry of its most stable isomer, which leads to the largest photoionization delays (up to 425~as) found in our calculations.

Interestingly, a further increase of the cluster size does not increase the delays further. The reason is obvious from Fig. 3d. Most larger clusters than the tetramer have a lower symmetry, many of them having no symmetry elements at all. This leads to a localization of the orbital densities on a small number of typically 2-3 neighboring molecules. This effect is reflected as a stagnation of the associated photoionization delays around values of $\sim$300~as in SB12. This observation suggests that the disorder-induced orbital localization in the larger clusters causes the experimentally observed saturation of the measured photoionization delays at the largest cluster sizes measured in this work.

\begin{figure}
    \centering
    \includegraphics[width=1.0\textwidth]{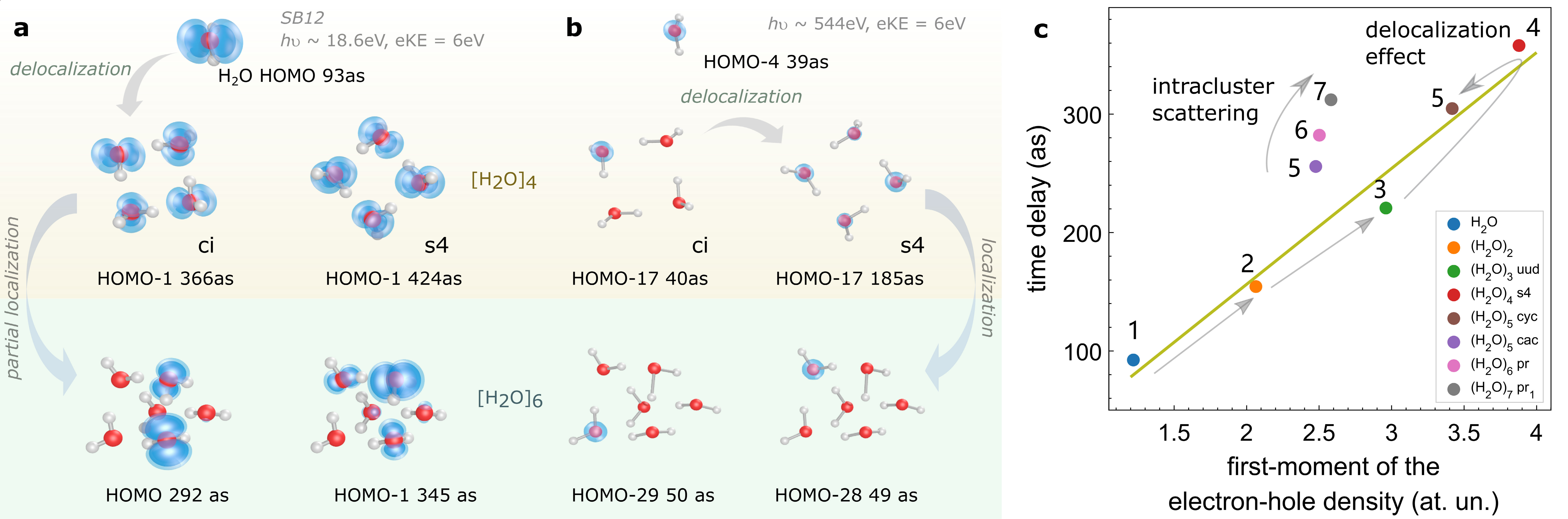}
    \caption{
    \textbf{Effect of orbital delocalization on photoionization delays of water clusters} \textbf{a}, Photoionization delays for the 1b$_1$ band of water clusters for a kinetic energy of eKE = 6.0~eV, corresponding to the experimental measurements in SB12. \textbf{b}, Same as \textbf{a}, but for the O1s-band of water clusters using the same kinetic energy, and a correspondingly adjusted photon energy. \textbf{c}, Correlation between the photoionization time delays and the first moment of the electron-hole density of the 1b$_1$ band of water clusters (eKE = 6.0~eV).
    }
\end{figure}

To further verify this surprisingly simple relation between time delays and orbital localization, we performed additional calculations on the oxygen-1s orbitals of the water clusters, with complete results shown in the SM (Figs. S14-S19). The O1s-orbitals have the advantage of remaining essentially atomic in character and not being significantly modified by hydrogen bonding and orbital hybridization. For this reason, they allow us to isolate the effect of orbital (de)localization even more clearly. The results obtained for the O1s-band are shown in Fig. 4b, where they are compared to the results for the 1b$_1$ band (Fig. 4a). This comparison further supports our interpretation. The photoionization delay remains essentially unchanged from the monomer (39~as) to the tetramer ci (40~as), where the O-1s orbital is fully localized, but increases to 185~as in the tetramer s4, where the O-1s orbital is fully delocalized as a consequence of the high symmetry. Increasing the cluster size further results in a complete localization of the O-1s orbital, which leads to a remarkable decrease of the photoemission delay to 49-50~as in the hexamer. The photoionization delays of the O1s orbitals in the hexamer being practically identical with the monomer (39~as) highlights, both, the direct link between photoionization delay and orbital delocalization and the very small effect (a few attoseconds) of electron scattering on the neutral H$_2$O moieties of the clusters.

Finally, we establish a quantitative link between time delays and electron-hole delocalization. Figure 4c shows a correlation plot between photoionization delays of the 1b$_1$ band of the water clusters (H$_2$O)$_n$ and the first moment of the electron-hole density in the 1b$_1$ band of the singly-ionized clusters (H$_2$O)$_n^+$ (defined in Eqs. (2)-(3) in Methods). The most stable isomers of all clusters from $n=1$ to $n=5$ display a nearly perfect linear relationship between the two quantities, as indicated by the straight line in Fig. 4c (correlation coefficient $\rho=0.984$). This representation also highlights the continuous increase of the delocalization from $n=1$ to $n=4$, followed by partial localization at $n=5-7$, visible in the fact that the second-most stable isomer of $n=5$ and the most stable isomers of $n=6$ and 7 have nearly the same first moment of the electron-hole density. Interestingly, the delays of the latter three isomers display a continuous, yet very small increase, which we assign to intracluster scattering, quantifying this effect as well.

These results have a range of interesting implications. First, they demonstrate the sensitivity of photoionization delays to orbital delocalization in water clusters, which, to our knowledge, has not been experimentally accessible in any form of matter so far. Second, they reveal the mechanism that is responsible for orbital localization in liquid water on the molecular level, i.e. the onset of structural disorder. This effect is reminiscent of Anderson localization in solids\cite{anderson1958}. Whereas perfect crystals with translational symmetry have fully delocalized bands, the presence of defects causes their localization, which has a multitude of interesting consequences in solid-state physics. The orbital localization in liquid water can thus be viewed as a consequence of its structural disorder.

We have introduced a new technique, ASCS, and have applied it to measure photoionization time delays of size-resolved water clusters. This study has revealed an unexpectedly simple relationship between orbital localization and time delays, establishing an experimental pathway to probing electron-hole localization in complex matter. Looking forward, our methods can be used to temporally resolve both local and non-local electronic relaxation dynamics in size-resolved water clusters, such as Auger decay, intermolecular Coulombic decay\cite{jahnke10a,mucke10a} and electron-transfer-mediated decay\cite{unger17a}. More generally, they will facilitate a molecular-level understanding of attosecond electron dynamics in the liquid phase, with implications for the elementary processes underlying chemical reactivity and biological function.

\section*{Methods}
\subsection*{Laser Setup and Attosecond-Pulse Generation}
The experimental setup is based on a regeneratively amplified Titanium-Sapphire laser system which delivers near-infrared (IR) femtosecond laser pulses with 1.2 mJ energy at 5 kHz repetition rate, a central wavelength of 800 nm and 36 fs pulse duration (full-width at half-maximum in intensity). This laser beam is split with a 70:30 beam splitter and the more intense part is focused into a 3 mm long, xenon-filled gas cell to generate an extreme-ultraviolet attosecond pulse train (XUV-APT) via high-harmonic generation. A coaxial 100-nm aluminum foil on a quartz ring is placed before a nickel-coated toroidal mirror (\emph{f} = 50 cm) to spectrally filter the XUV spectrum and eliminate the residual IR pulse co-propagating with the XUV beam. The XUV spectrum was characterized with a home-built XUV spectrometer consisting of a 100-nm aluminum film, an aberration-corrected flat-field grating (Shimadzu 1200 lines/mm) and a micro-channel-plate (MCP) detector coupled to a phosphor screen. The XUV-APT is recombined  with the remaining part of the IR beam after the toroidal mirror via a perforated silver mirror to constitute a Mach-Zehnder interferometer. The path length difference, i.e. the time delay between the overlapping XUV-APT and IR pulses is controlled via a high-precision direct-current motor (PI, resolution 0.1 $\mu$m) and a piezoelectric motor (PI, resolution 0.1 nm), constituting a combined delay stage operating on femtosecond and attosecond time scales, respectively.

\subsection*{Coincidence Spectrometer}
The phase-locked XUV-APT and IR pulses are collinearly focused into the supersonic gas jet in a COLTRIMS (COLd Target Recoil Ion Momentum Spectroscopy)\cite{Dorner2000a,Jagutzki2002,Ullrich_2003} spectrometer. The electrons and ions created by XUV photoionization  are guided by a weak homogeneous electric field (3.20 V$\cdot$cm$^{-1}$) and a homogeneous magnetic field (6.70~G) towards two time- and position-sensitive detectors at opposite ends of the spectrometer. The detectors consist of two MCPs (Photonis) in Chevron configuration, followed by a three-layer delay-line anode (HEX) with a crossing angle of 60 degrees between adjacent layers and an active radius of 40 mm manufactured by RoentDek. For the electrons, the length of the extraction region is 7~cm, followed by a 14~cm field-free region. A homogeneous magnetic field is applied over all regions by a set of Helmholtz coils, which are tilted to counteract the earth's magnetic field. The COLTRIMS gives access to the typical electron-ion coincidence measurement with full three-dimensional momentum resolution in 4$\pi$ solid angle. The momentum resolution of electrons is $\Delta p_{x}$ = $\Delta p_{y}$ = 0.001 a.u. and $\Delta p_{z}$ = 0.056 a.u., where $x$ corresponds to the direction of light propagation, $y$ is the direction of the supersonic gas jet and $z$ is the time-of-flight direction. The photoelectron kinetic energy is calibrated via the XUV-APT photoelectron spectrum of argon with an ionization potential of $I_p \sim$ 15.8 eV.

\subsection*{Cluster source}
The neutral water clusters\cite{Hagena1972,pradzynski2012} are formed in a continuous supersonic expansion into vacuum with a water vapor pressure of 0.3 MPa through a 50 $\mu$m nozzle orifice and pass through two conical skimmers (Beam Dynamics) located 10 mm and 30 mm downstream with a diameter of 200 $\mu$m and 1~mm, respectively. The liquid water is maintained at 408~K to give rise to a sufficient vapor pressure in a container of 0.7~L to support a stable water-cluster beam for a duration of 120 hours. The water-cluster source is coupled to the COLTRIMS via two differential pumping stages. To maintain the ultrahigh vacuum in the main reaction chamber, a differentially pumped beam dump captures the molecular beam after the interaction region.

\subsection*{Definition of the calculated quantities}

The calculation of the orbital-specific photoionization delays is described in the main text and in more detail in Section 2 of the SM. Here, we additionally define the cross-section-averaged delays shown in Figs.~3a, 3b as large empty circles and also in Fig.~4c. Since the contributions of individual orbitals to the 1b$_1$ band of the water clusters cannot be resolved, we introduce the cross-section-average of the time delays over the $n$ orbitals constituting the 1b$_1$ band of (H$_2$O)$_n$ as follows:
\begin{equation}
    \tau(E)=\frac{\sum_{i=1}^n \sigma_i(E)\tau_i(E)}{\sum_{i=1}^n \sigma_i(E)},
\end{equation}
where $\sigma_i(E)$ is the photoionization cross section of orbital $i$ of the 1b$_1$ band at the photon energy $E$ and $\tau_i(E)$ is the corresponding photoionization time delay.

In Fig.~4c, we additionally show the first moment of the electron-hole density, 
calculated using ORBKIT package \cite{hermann2016orbkit}. 
In the case of a single orbital (with index $i$) this is defined as
\begin{equation}
    \mathcal{M}_i=\frac{\int\rho_i(\vec{r}-\vec{r}_i)\left|\vec{r}-\vec{r}_i\right|d^3r}{\int\rho_i(\vec{r})d^3r},
\end{equation}
where $\rho_i(\vec{r})$ is the density of orbital $i$ and $\vec{r}_i$ is its center of charge.
In analogy to the time delays, we also define a cross-section average of the orbital delocalization as
\begin{equation}
    \mathcal{M}=\frac{\sum_{i=1}^n \sigma_i\mathcal{M}_i}{\sum_{i=1}^n \sigma_i},
\end{equation}

\subsection*{Acknowledgment}
 We thank A. Schneider and M. Seiler for their technical support.
{\bf Funding} We gratefully acknowledge funding from an ERC Consolidator Grant (Project No. 772797-ATTOLIQ), project 200021\_172946 as well as the NCCR-MUST, funding instruments of the Swiss National Science Foundation. D. J. thanks the European Union's Horizon 2020 programme (FP-RESOMUS - MSCA 801459) program for a fellowship and A. Schild for introduction to ORBKIT. The results have been obtained on the ETH Z\"{u}rich Euler cluster and the NCCR-Cluster supercomputer. \textbf{Author contributions} X.G. and S.H. carried out the experiments and analysed the experimental data. X.G. constructed the experimental apparatus with contributions from S.H., K.Z. and C.P..  D.J. and H.J.W. performed the theoretical calculations. X.G., S.H., and H.J.W wrote the initial manuscript. All authors discussed and reviewed the manuscript. \textbf{Competing Interests} The authors declare no competing interests.

\subsection*{Additional information}
 \textbf{Supplementary information} is available for this paper.\\
 \textbf{Correspondence and requests for materials} should be addressed to H.J.W.\\
 \textbf{Data availability statement} All data is available  from the corresponding author on request.



\newpage
\bibliography{attoH2On,attobib}%
\bibliographystyle{naturemag_noURL}

\end{document}


\baselineskip24pt

\maketitle 

\textbf{Contents}\\
\textbf{1. \quad Attosecond size-resolved cluster spectroscopy}\\
1.1 \quad Water-Cluster Generation setup\\
1.2 \quad Water-Cluster Fragmentation Dynamics and Abundance\\
1.3 \quad Water-Cluster Structure\\
1.4 \quad Attosecond Coincidence Interferometry\\
1.5 \quad Phase Reconstruction \\
\textbf{2. \quad Calculations}\\
2.1 \quad Calculations of photoionization matrix elements and cross sections\\
2.2 \quad Calculations of photoionization time delays\\
2.3 \quad Photoionization delays of the 1b$_1$ and O-1s (1a$_1$) bands of water clusters\\
2.4 \quad Effects of nuclear motion and resonances on photoionization delays of water clusters\\
2.5 \quad Benchmarking the quantum-scattering calculations against expeirmental photoelectron asymmetry parameters
\baselineskip26pt 

\newpage
\section*{\Large \textbf{1. Attosecond size-resolved cluster spectroscopy}}
\subsection*{1.1 Water-Cluster Generation setup}
The experimental setup and the water-cluster source are illustrated in Fig. S1. The neutral water clusters\cite{Hagena1972,pradzynski2012} were formed in a continuous supersonic expansion into vacuum with a water-vapor pressure of 0.3 MPa through a 50 $\mu$m nozzle orifice and passed through two conical skimmers (Beam Dynamics) located 10 mm and 30 mm downstream with a diameter of 200 $\mu$m and 1~mm, respectively. The liquid water was maintained at 408~K to give rise to a sufficient vapor pressure in a container of 0.7~L to support a stable water-cluster beam for a duration of 120 hours. The water-cluster source was coupled to the COLTRIMS via two differential pumping stages. To maintain the ultrahigh vacuum in the main reaction chamber, a differentially pumped beam dump captured the molecular beam after the interaction region.

\begin{figure}
\includegraphics[width=1.0\textwidth]{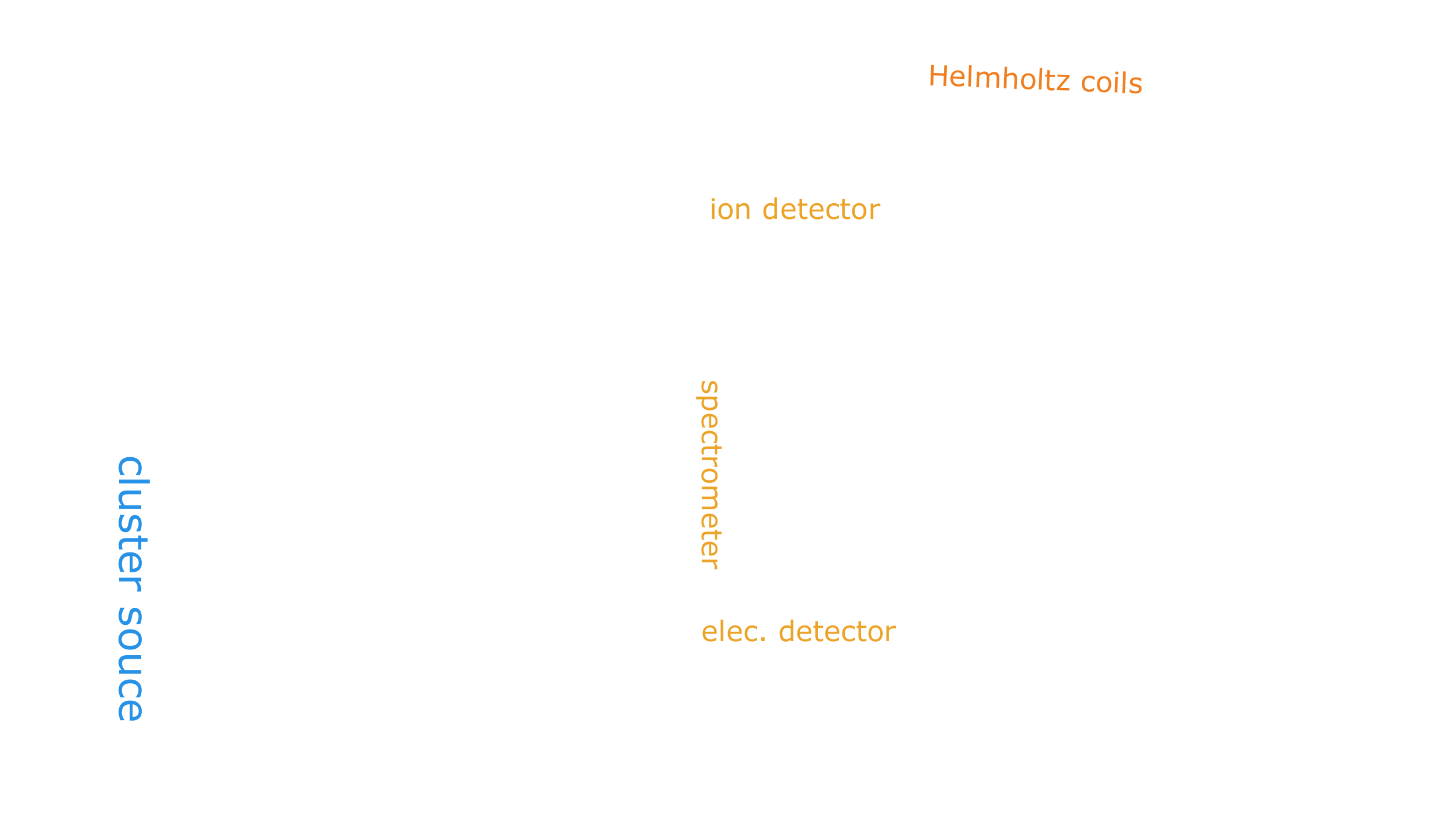}
\newline
{\textbf{Fig. S1: Drawing of the coincidence spectrometer and water-cluster source.} The water-cluster source is coupled to the electron-ion coincidence spectrometer through a two-stage differential pumping system with a molecular-beam dump on the opposite side. The time-of-flight axis of the coincidence spectrometer was perpendicular to the supersonic beam. Homogeneous electric fields were created by a stack of electrodes that guide the motion of the ions and electrons to the MCP detectors. The pair of Helmholtz coils provided the required magnetic fields.}
\label{setup}
\end{figure}

\subsection*{1.2 Water-Cluster Fragmentation Dynamics and Abundance}
The fragmentation dynamics of water clusters upon photoionization consists out of a rapid proton transfer, followed by the loss of a single neutral OH unit\cite{Dong2006,SHIROMARU1987,Shi1993,Wei1994,Bobbert2002,Tachikawa2004,Belau2007,Liu2011,Zamith2012,Litman2013,Uwe2013}.
\begin{equation}
    (\mathrm{H_2O})_n +h\nu \longrightarrow [(\mathrm{H_2O})^+_n]^* + e^- \longrightarrow (\mathrm{H_2O})_{n-1}\mathrm{H}^+ + \mathrm{OH} + \mathrm{e}^- .
\tag{S1}
\end{equation}
This pathway dominates the fragmentation of water clusters up to n=15\cite{Dong2006,Belau2007,Shi1993}. For larger clusters the subsequent loss of additional water molecules becomes relevant following the reaction
\begin{equation}
    (\mathrm{H_2O})_n+h\nu \longrightarrow (\mathrm{H_2O})_n^+ +e^- \longrightarrow(\mathrm{H_2O})_{n-1-m}\mathrm{H}^+ +\mathrm{OH} +\mathrm{e}^- + m\mathrm{H_2O}.
\tag{S2}    
\end{equation}
However, in photoionization experiments with 12.5 eV\cite{Belau2007} and 26.5 eV\cite{Dong2006}, the onset of $\mathrm{H_2O}$ molecule loss has only been observed for clusters with n $\geq$ 5, with the decay fraction for n=5 and n=6 still well below 10$\%$. Figure S2 shows the measured intensity distribution of the protonated water clusters $(\mathrm{H_2O})_n\mathrm{H}^+$ of our cluster source. The vanishing intensity for larger clusters allows us to exclude the possibility of fragmentation dynamics from large clusters.

\begin{figure} 
\begin{center}
\includegraphics[width=0.5\textwidth]{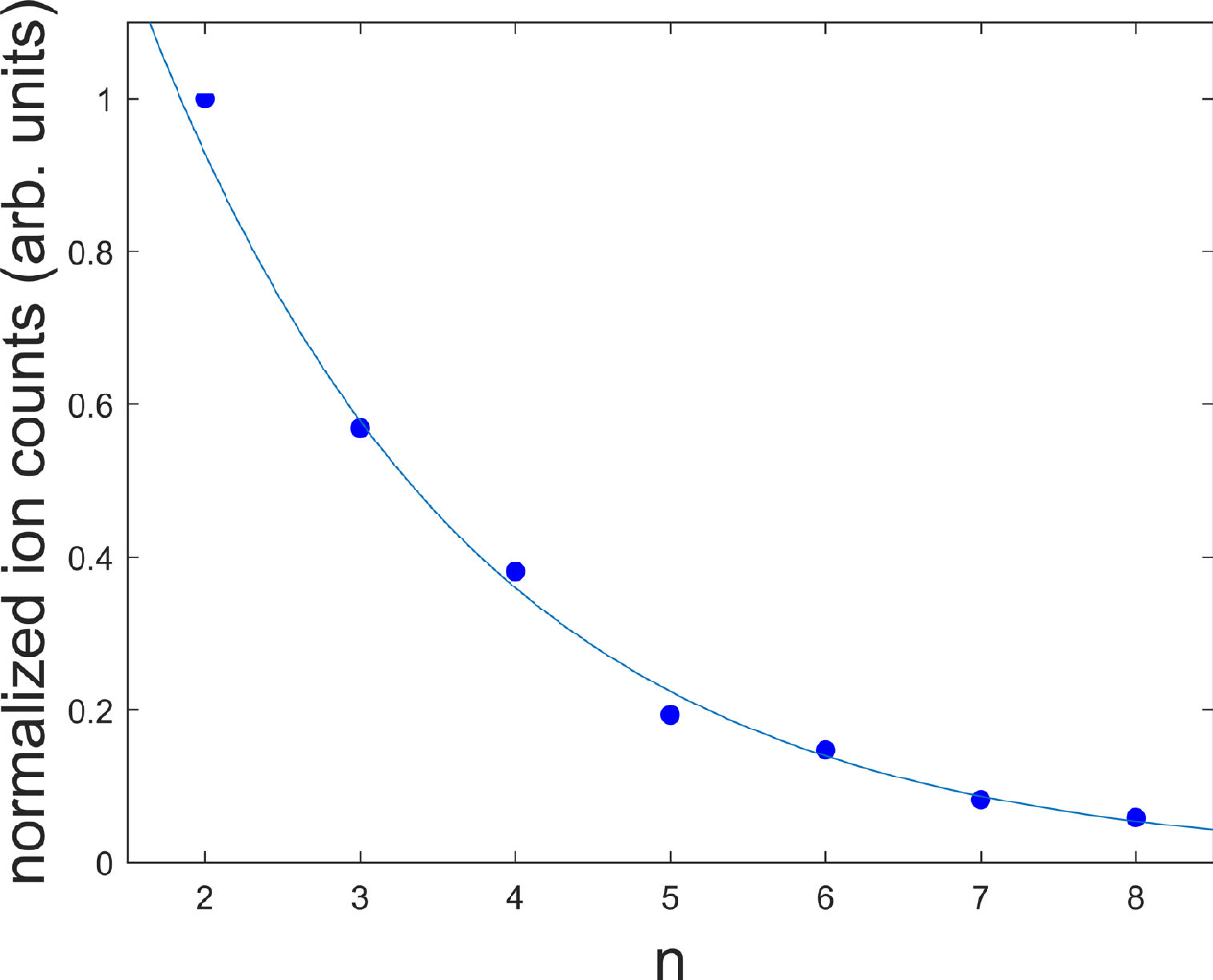}
\end{center}
{\textbf{Fig. S2: Measured intensity of the protonated water-cluster ions $(\mathrm{H_2O})_n\mathrm{H}^+$ in our cluster beam.} The data was normalized with respect to n=2, i.e. $(\mathrm{H_2O})_2\mathrm{H}^+$. The blue solid line is an exponential fit to the experimental data (blue dots).}
\label{h2onyield}
\end{figure}

\subsection*{1.3 Water-Cluster Structure}

This section summarizes the isomer structures of small water clusters and their population distribution within the temperature range of our cluster beam. The structures of small water clusters and their respective energies have been studied extensively in recent years\cite{shi93h2on,Temelso2011a,Malloum2019,Liu1996,Liu1996a,Xantheas2000,Richardson2016,hartweg2017a,cvitas20a}. Table S1 shows the electronic binding energies $\Delta E_e$ of the clusters with respect to the (non-interacting) monomers and the enthalpies of formation $\Delta H(0K)$ at 0~K
for different isomers as calculated by Temelso et al.\cite{Temelso2011a}. For an estimated internal temperature of the water clusters of 150~K there is 0.3~kcal/mol of thermal energy available. Therefore, we only take into account the isomeric structures that lie within 0.3 kcal/mol in enthalpy of the energetically most stable isomer. The resulting isomers are identified with bold font in Table S1.

\begin{table}
{Table S1: Binding energy and enthalpy of formation of different water cluster isomers given in kcal/mol. 
The values were taken from  Ref.\cite{Temelso2011a}.}
\begin{small}
\begin{center}
\begin{tabular}{lccc}
\hline \hline
   \multicolumn{1}{c}{(H$_2$O)$_n$}   &
   \multicolumn{1}{c}{$\Delta E_e$}   &
   \multicolumn{1}{c}{$\Delta H(0K)$} 
   \\
   \hline
    \textbf{2-(C$_s$)}    &-5.03    &-3.14   \\ 
    \textbf{3-UUD}        &-15.70   &-10.63  \\ 
    \textbf{3-UUU(C$_3$)} &-15.08   &-10.40  \\ 
    \textbf{4-(S$_4$)}    &-27.43   &-19.74  \\ 
     4-(C$_i$)            &-26.58   &-19.08  \\ 
     4-PY                 &-23.88   &-16.56  \\ 
     \textbf{5-CYC}       &-36.01   &-26.29  \\ 
     5-FR-B               &-34.87   &-25.11  \\ 
     5-CA-C               &-34.69   &-24.70  \\ 
     5-CA-A               &-34.54   & -24.53  \\ 
     5-CA-B               &-33.82   & -23.83  \\ 
     5-FR-C               &-32.44   & -22.87  \\ 
     5-FR-A               &-33.13   & -23.23  \\ 
    \textbf{6-PR}         & -46.14  & -33.16  \\ 
    \textbf{6-CA}         & -45.93  & -33.14  \\ 
     6-BK-1               & -45.51  & -33.11  \\ 
     6-BK-2               & -45.14  & -32.74  \\ 
     6-CC(S$_6$)          & -44.60  & -32.76  \\ 
     6-BAG                &  -44.59 & -32.36  \\ 
     6-CB-1(C$_2$)        & -43.57  & -31.91  \\ 
     6-CB-2(C$_2$)        & -43.51  & -32.03  \\ 
\hline
\hline
\end{tabular}
\label{tab:clusters}
\end{center}
\end{small}
\end{table}

\subsection*{1.4 Attosecond Coincidence Interferometry}

In our experiments, photoionization of water clusters was induced by single-XUV-photon absorption, yielding a main-band (MB) photoelectron band at a kinetic energy $eKE = E_{XUV} - I_{p}$, as shown in Fig. S3. The synchronized IR field induced two distinct quantum paths from adjacent main bands to the same final side-band (SB) state by absorption or stimulated emission of one IR photon. The interference between these two quantum paths gave rise to the typical SB intensity oscillations with half the IR period as a function of the APT-IR pump-probe delay. To achieve attosecond temporal stability, a frequency-stabilized laser beam (He-Ne continuum reference laser, $\lambda$ = 632.8nm) was coupled into the XUV-APT and IR Mach-Zehnder interferometer\cite{Chini2009,Huppert2015,Gong2017}, resulting in a time jitter of $\sim$ 40 as, as shown in Fig. S4.

The photoionization time delays of water clusters are determined from the intensity oscillations of the SB peaks defined as
\begin{equation}
    A_q(\tau) = A\cos(2\omega(\tau - \tau_q))+B = A\cos(2\omega(\tau - \tau_q^{\rm XUV} - \tau_q^{\rm sys}))+B,
\tag{S3}        
\end{equation}
where $\omega$ is the angular frequency of the IR field, $\tau_q^{\rm XUV}$ represents the emission-time difference of harmonics $q+1$ and $q-1$, i.e. it reflects the attochirp of the XUV-APT\cite{paul2001rb} and $\tau_q^{\rm sys}$ is the system-specific photoionization time delay\cite{wigner1955,Smith1960,kluender11a,dahlstrom12a,huppert16a,baykusheva17a}.

\begin{figure}
\begin{center}
\includegraphics[width=0.8\textwidth]{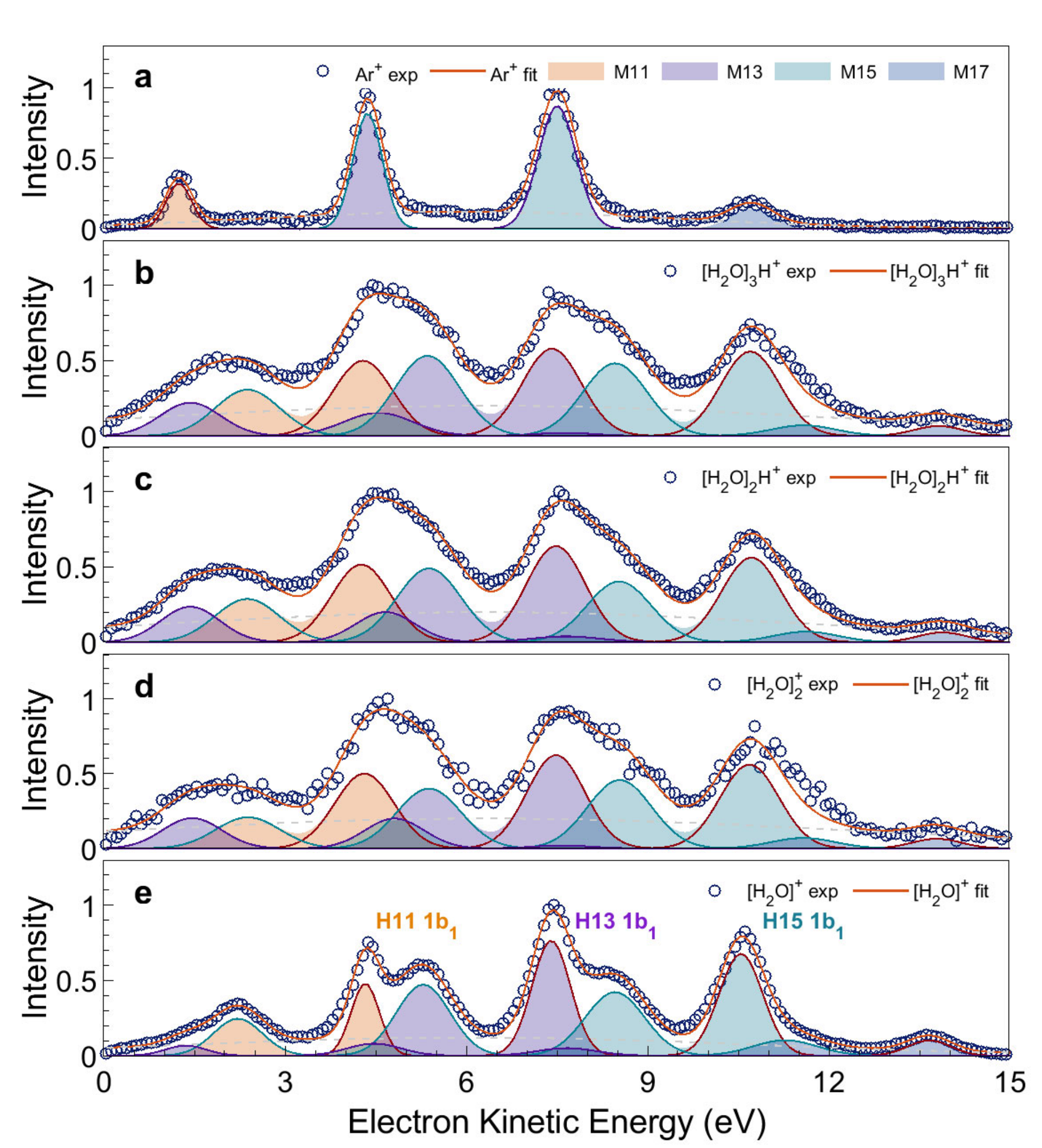}
\end{center}
{\textbf{Fig. S3: XUV-only photoelectron spectra of water clusters and argon atoms.} 
\textbf{a} argon, \textbf{b} [H$_2$O]$_3$H$^+$, \textbf{c} [H$_2$O]$_2$H$^+$, \textbf{d} [H$_2$O]$_2^+$, \textbf{e} [H$_2$O]$^+$. The XUV-APT included the high-harmonic orders from 11 to 17.}
\label{h2onPES}
\end{figure}

\begin{figure}
\begin{center}
\includegraphics[width=0.5\textwidth]{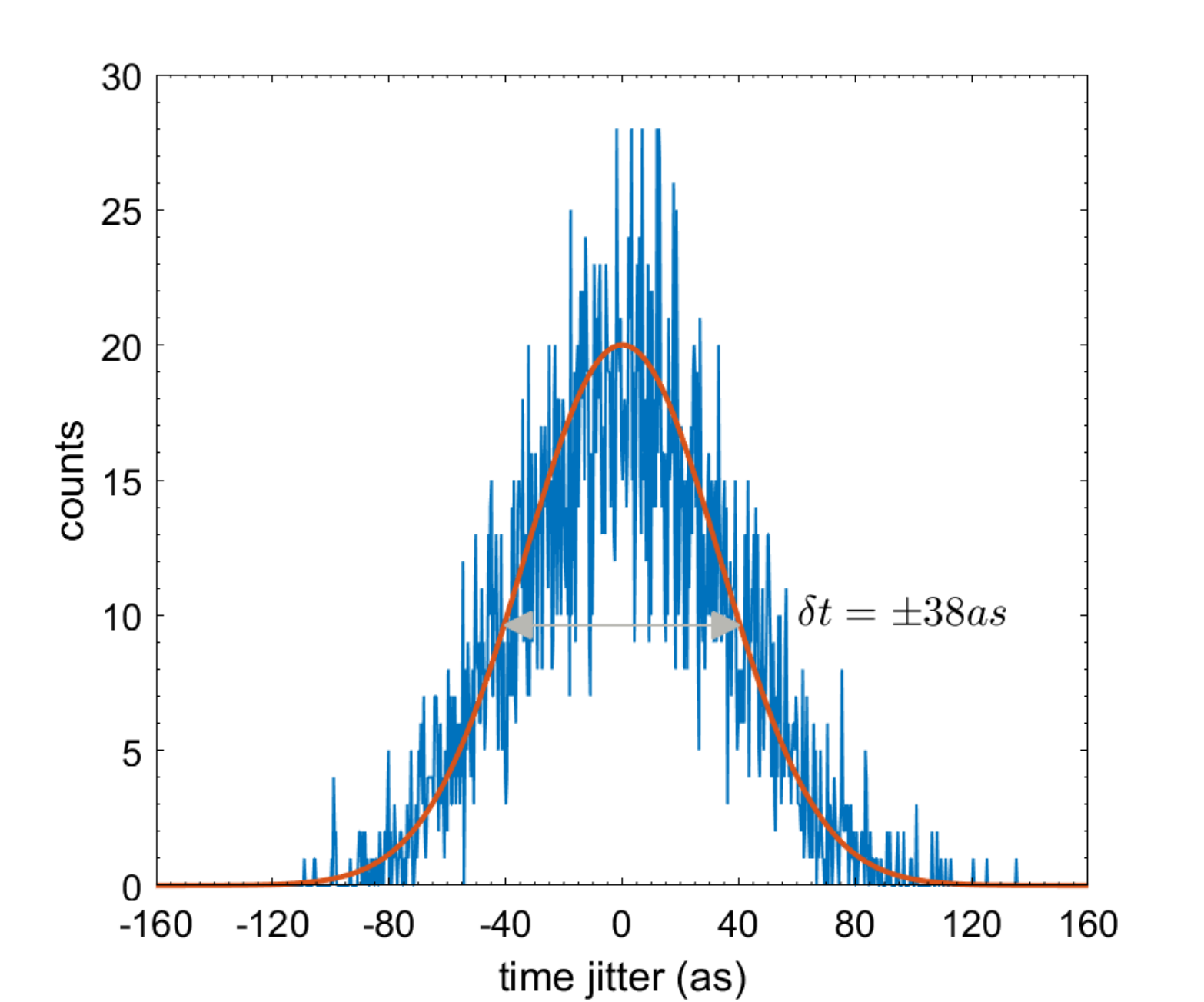}
\end{center}
{\textbf{Fig. S4: Delay stability of the attosecond pump-probe interferometer.} The time jitter of the phase-locked Mach-Zehnder interferometer amounted to $\pm 38$as.}
\label{jitter}
\end{figure}

\subsection*{1.5 Phase Reconstruction}

Compared to atomic systems, the measurement of molecular photoionization time delays typically faces the challenge of spectral overlap\cite{huppert16a,jordan18b}. This is because the XUV-APT ionizes electrons out of several possible orbitals, which leads to spectral overlap between the photoelectron MB spectra created by different harmonic orders and the SB spectra. Figure S5 shows the attosecond photoelectron spectra recorded in coincidence with different water-cluster fragments, ranging  from the dimer to the protonated pentamer. To extract the photoionization time delays from spectrally overlapping attosecond photoelectron spectra, we used the complex-valued principal components analysis as described in detail in\cite{jordan18b}, which has also been successfully used in our recent work on liquid water\cite{jordan20a}. In the first step, the XUV-only photoelectron spectrum is fitted with a set of Gaussians, as shown in Fig. S3. In a second step, we fitted additional Gaussians to reproduce the XUV+IR photoelectron spectrum. Next, a Fast-Fourier Transformation (FFT) was done line by line on the attosecond photoelectron spectra (Fig. S5) along the time-delay axis and the resulting band in the complex-valued FFT at the 2$\omega$ angular frequency was fitted by multiplying each Gaussian component obtained in the XUV+IR fit with a complex amplitude $e^{z_j}$
\begin{equation}
    I_{fit}(E) = \sum_{j}p_j(E)e^{z_j} = \sum_{j}\underbrace{e^{a_j}p_j(E)}_{A_j(E)}e^{ib_j},
\tag{S4}
\end{equation}
where $p_j(E)$ is the Gaussian fit for photoelectron band $j$ (see Fig. S3). This complex number $z_j=a_j+ib_j$ simultaneously accounts for the side-band specific delay $\tau_j=-b_j/(2\omega)$ and a finite modulation contrast when $\left|e^{a_j}\right|<1$. The numerical robustness of the obtained phase shifts of SB12 and SB14 is illustrated in Fig. S6, obtained by randomly varying the initial guesses of the 1b$_1$ binding energies over a range of 0.2~eV. These results showed that the determination of the phase shifts was robust with respect to the initial guesses and demonstrated the absence of any systematic correlation between these two quantities.

To exclude any major contributions due to vibrational effects in the measured photoionization time delays, we additionally performed the attosecond size-resolved measurement with pure D$_2$O clusters (Sigma-Aldrich, 99.9 atom $\%$ D). Figure S7 shows the measured mass spectrum of ionized D$_2$O clusters, under similar conditions as the spectrum shown in Fig. 1\textbf{c} and 1\textbf{d} in the main text. The ion mass spectra for the different isotopes look very similar, as the fragmentation pathways are the same \cite{Radi1999,Huang2016}. Figure S8 shows the size-resolved attosecond photoelectron spectra and their numerical complex-fitting results for the heavy water clusters. 

\begin{figure}
\begin{center}
\includegraphics[width=1.0\textwidth]{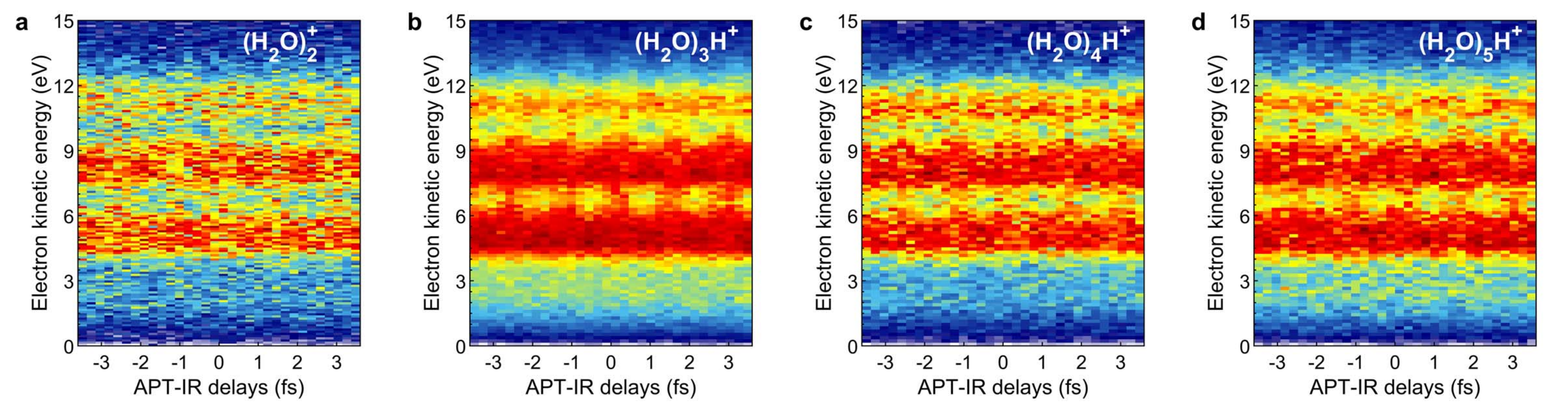}
\end{center}
{\textbf{Fig. S5: Attosecond photoelectron spectra for different sized water clusters.} \textbf{a}, Attosecond photoelectron spectra coincident with the unprotonated water dimer cation, (H$_2$O)$_2^+$. \textbf{b}-\textbf{d}, Same as \textbf{a} but for \textbf{b} (H$_2$O)$_3$H$^+$, \textbf{c} (H$_2$O)$_4$H$^+$, \textbf{d} (H$_2$O)$_5$H$^+$.}
\label{H2OnRB}
\end{figure}

\begin{figure}
\thispagestyle{empty}
\begin{center}
\includegraphics[width=0.85\textwidth]{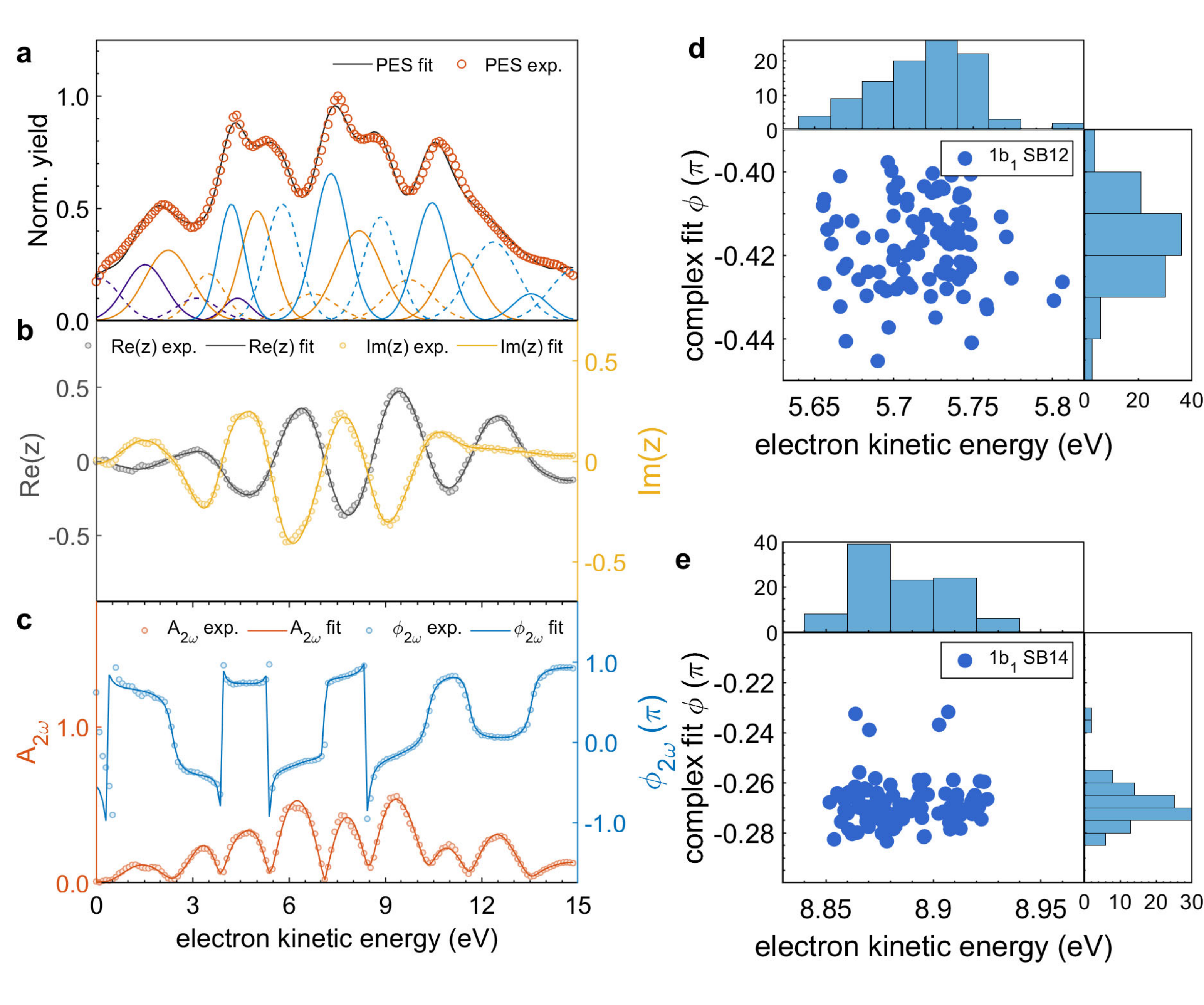}
\end{center}
{\textbf{Fig. S6: Numerical stability of the phase reconstruction} \textbf{a}, Measured (red circles) and fitted (black line) photoelectron spectrum, in coincidence with H$_2$O$^+$ and in the presence of the XUV-APT + IR field with the XUV single-photon contributed main bands (solid lines) and sidebands (dashed lines) corresponding to the $1b_1$ (blue), $3a_1$(orange) and $1b_2$ (purple) bands. \textbf{b}, Real (black) and imaginary (yellow) components of the 2$\omega$ frequency band in the Fourier transform of the RABBIT photoelectron spectrum of water monomers. The circles represent the data points and the full lines represent the complex fit. \textbf{c}, Same as \textbf{b}, showing the amplitude and phase of the 2$\omega$ frequency band of the Fourier transform. \textbf{d}, Retrieved SB12 phase of $1b_1$ band as a function of the initial guess of the electron-kinetic energy. The upper and right-hand panels show the projected electron-kinetic-energy and phase distributions. \textbf{e}, Same analysis as in d for the phase of SB14.}
\label{cfit}
\end{figure}

\begin{figure}
\begin{center}
\includegraphics[width=0.8\textwidth]{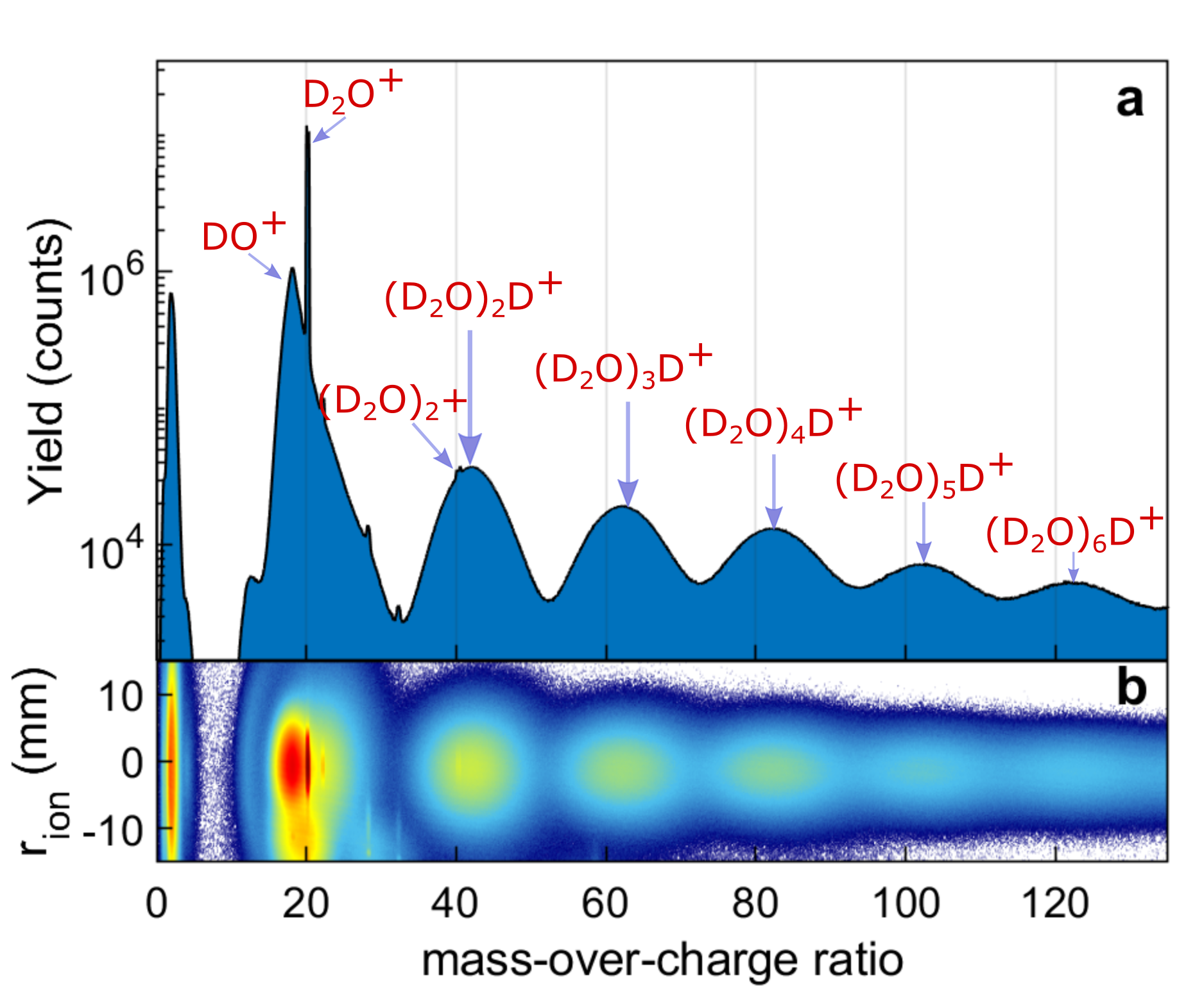}
\end{center}
{\textbf{Fig. S7: Mass spectrum of D$_2$O clusters.} 
\textbf{a}, Mass spectrum of the cluster beam photoionized by an APT as a function of the mass-over-charge (MOC) ratio.
\textbf{b} Two-dimensional MOC spectrum of water cluster species as a function of position on the ion detector.}
\label{d2omq}
\end{figure}

\begin{figure}
\begin{center}
\includegraphics[width=0.9\textwidth]{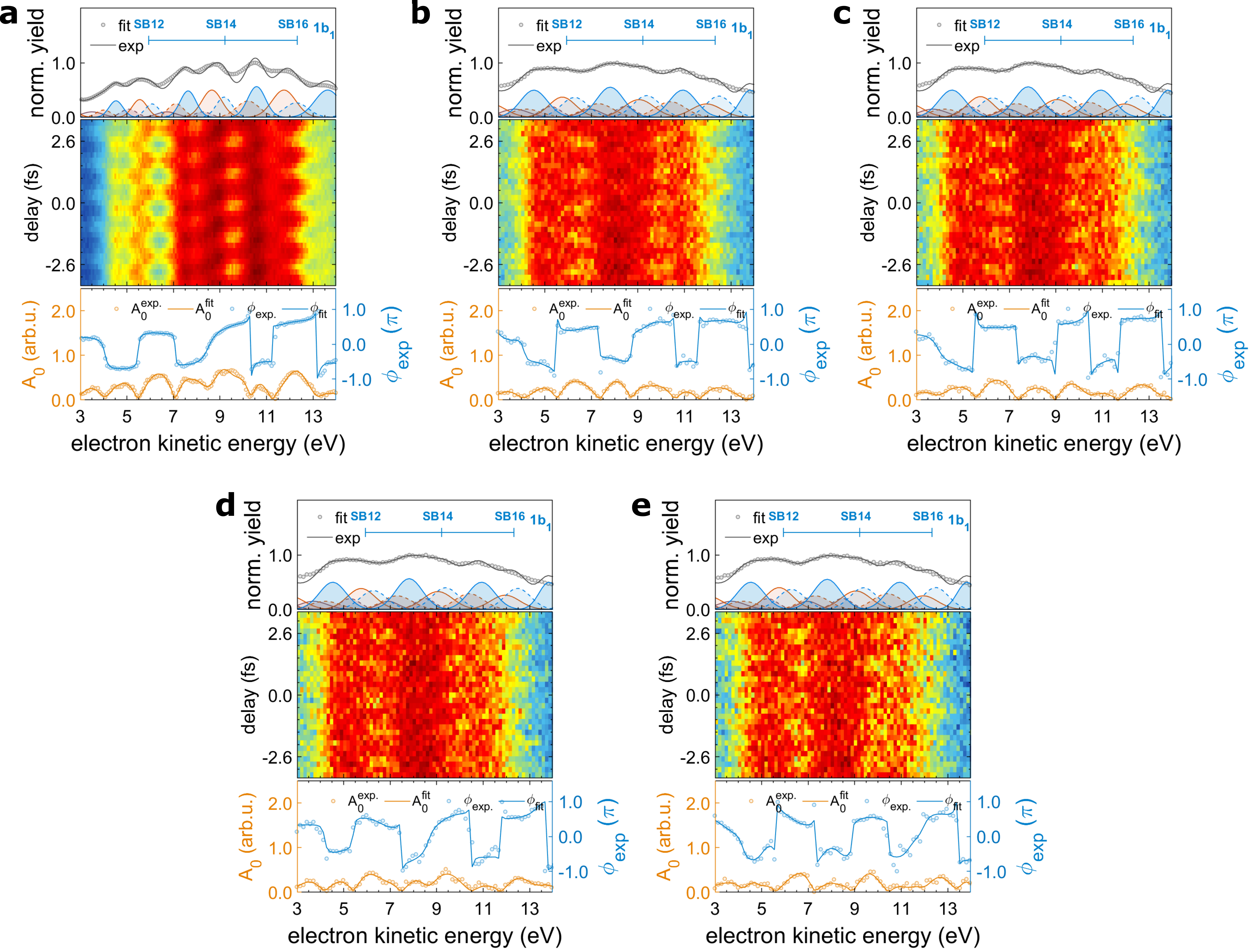}
\end{center}
{\textbf{Fig. S8: Attosecond photoelectron spectroscopy of size-selected D$_2$O clusters.} 
Attosecond photoelectron spectra created by overlapping XUV-APT and IR pulses and detected in coincidence
with \textbf{a} $\mathrm{D}_{2}\mathrm{O}^{+}$, 
\textbf{b} $(\mathrm{D}_{2}\mathrm{O})_{2}\mathrm{D}^{+}$, 
\textbf{c} $(\mathrm{D}_{2}\mathrm{O})_{3}\mathrm{D}^{+}$, 
\textbf{d} $(\mathrm{D}_{2}\mathrm{O})_{4}\mathrm{D}^{+}$, 
\textbf{e} $(\mathrm{D}_{2}\mathrm{O})_{5}\mathrm{D}^{+}$, respectively. The spectrum shown in upper panel is integrated over the APT-IR delay, the middle panel shows the RABBITT trace, and the bottom panel displays the Fourier transforms at 2$\omega$ of the attosecond photoelectron spectra, shown in terms of their modulation amplitude (orange color) and phase (blue color). The experimental data and the fitted curves are shown as open circles and solid lines, respectively.}
\label{D2OnRB}
\end{figure}

\clearpage
\section*{\Large \textbf{2 Calculations}}
\subsection*{2.1 Calculations of photoionization matrix elements and cross sections}

To describe the photoionization dynamics of an $N$-electron molecule or cluster, we employed a quantum-scattering calculation based on the iterative Schwinger variational principle\cite{Lucchese1984}. In this approach, we performed a single-center partial-wave decomposition of the initial electronic state, $\Psi_{i} = \sum_{\ell m}R_{n\ell}(r)Y_{\ell m}(\hat{r})$, which was constructed from an anti-symmetrized product of the $n = N/2$ occupied orbitals, as implemented in \emph{ePolyScat}\cite{Lucchese94,Lucchese99}. The final-state wave function and the scattering potential were also expanded into partial waves.

The photoionization dipole matrix elements in the length gauge for linearly polarized radiation are given by
\begin{equation}
 I_{i,f} = \left \langle \Psi_{f,\mathbf{\kappa}}^{(-)} \left | \mathbf{r}\cdot \hat{\mathbf{E}}_{XUV} \right | \Psi _{i}\right \rangle  = \sqrt{\frac{4\pi}{3 \mathbf{{\rm\kappa}}} }\sum_{\ell mv}I_{\ell mv}Y_{\ell m}(\mathbf{\hat{\kappa}}) Y_{1v}^{*}(\mathbf{\hat{n}})
\label{eps:dipolgxuv}
\tag{S5}
\end{equation}
where $\mathbf{r}$ is the position operator, $Y_{\ell m}$ are the spherical harmonics, $Y_{1,v=0,\pm1}$ describes the orientation $\hat{\mathbf{E}}_{XUV}$ of the XUV polarization in the molecular frame, $\bf{\hat{\kappa}}$ is the asymptotic momentum of the outgoing photoelectron wavepacket, $|\Psi_{f,\bf{\kappa}}^{(-)}\rangle $ denotes the observed final-state wave function, and  $I_{\ell mv} = \sqrt{\frac{2}{\pi}}(-i)^{\ell}\left \langle \Psi_{f,\mathbf{\kappa}\ell m}^{(-)} \left | \mathbf{r}_{v}\right | \Psi _{i}\right \rangle$ is the partial-wave matrix element.
The photoionization cross sections integrated over target-orientation and photoelectron-emission angles are given by
\begin{equation}
\sigma = \frac{4\pi^{2}E}{3c}\sum \left | I_{i,f} \right |^2,
\label{eps:cs}
\tag{S6}
\end{equation}
where $E$ is the photon energy.

\subsection*{2.2 Calculations of photoionization time delays}

As introduced by Wigner and Smith\cite{wigner1955,Smith1960}, the energy derivative of the scattering phase is associated with a time delay in scattering. In photoionization, this definition leads to a delay of the outgoing photoelectron wavepacket. The case of atoms and molecules have been treated in Refs.\cite{dahlstrom12a} and \cite{baykusheva17a}, respectively. The photoionization time delay is defined as the energy derivative of the complex photoionization amplitude $f(\epsilon)$
\begin{equation}
    \tau = \frac{d}{d\epsilon}{\rm Arg}(f(\epsilon))={\rm Im}\left\{ \frac{1}{f(\epsilon)}\frac{df}{d\epsilon} \right\}.
    \tag{S7}
\end{equation}
In the picture of partial-wave decomposition, the outgoing wavepacket is expressed as a coherent sum over partial waves, $\Psi = \sum_{lm}\psi_{lm}$, where each component is defined by the quantum numbers ($l,m$), i.e. the electronic orbital angular momentum and its projection onto a given quantization axis, and each ($l,m$) defines a partial-wave scattering channel.

In contrast to the spherical nature of the scattering potential in the atomic case, a molecule or cluster has a highly anisotropic potential, which requires many partial waves in the single-center expansion. In our calculations we have typically used $l_{\rm max}=50$.
The photoionization time delay in the molecular frame depends on the photon energy $E$, the final momentum  $\hat{\mathbf{\kappa}}$ and the polarization direction $\hat{\mathbf{E}}_{XUV}$ of the ionizing radiation. It can be defined as:
\begin{equation}
\tau_{\omega}^{MF}(E,\hat{\mathbf{\kappa}},\hat{\mathbf{E}}_{XUV}) = \hbar\frac{\partial}{\partial E}[{\rm Arg}(\left \langle \Psi_{f,\mathbf{\kappa}}^{(-)} \left | \mathbf{r}\cdot \hat{\mathbf{E}}_{XUV} \right | \Psi _{i}\right \rangle)].
\tag{S8}
\end{equation}

Figure S\ref{R9} illustrates the photoionization delay (left) and the differential photoionization cross section (right) for the case of the water dimer as a function of the emission angles ($\theta,\phi$) in the molecular frame.

\begin{figure}[h]
\begin{center}
\includegraphics[width=0.8\textwidth]{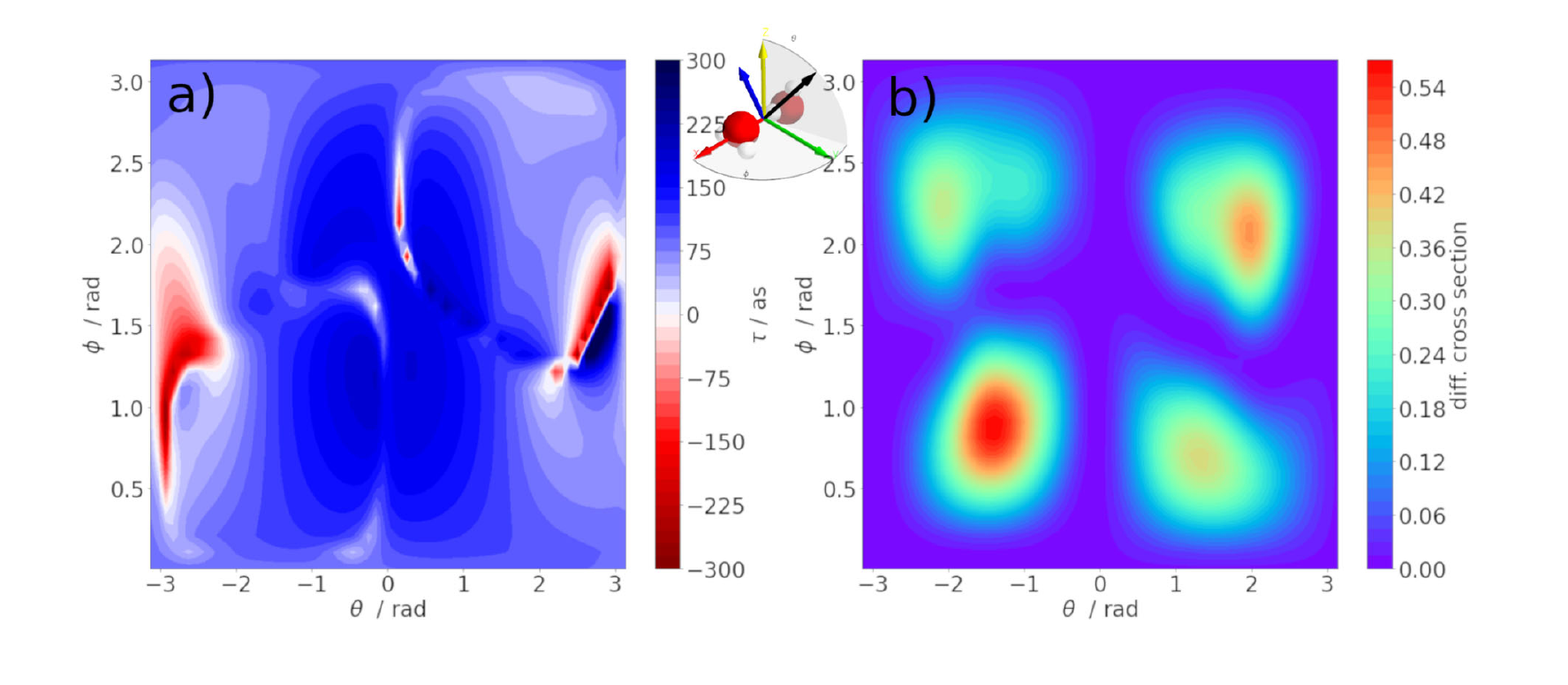}
\end{center}
\setcounter{figure}{8} 
\let\nobreakspace\relax
\captionsetup{labelfont=bf}
\renewcommand\figurename{\textbf{Fig. S}}
\def\fnum@figure{\figurename\thefigure}
\caption{\textbf{Example of one-photon (Wigner) time delay in the molecular frame (a) and differential photoionization cross section (b)} for HOMO of the (H$_2$O)$_2$ system and the light polarisation direction (blue arrow) indicated in the inset, as a function of photoelectron ejection directions (black arrow defined by $\theta$ and $\phi$ angles in molecular frame).}
\label{R9}
\end{figure}

\begin{figure}[h]
\centering
\includegraphics[width=0.7\textwidth]{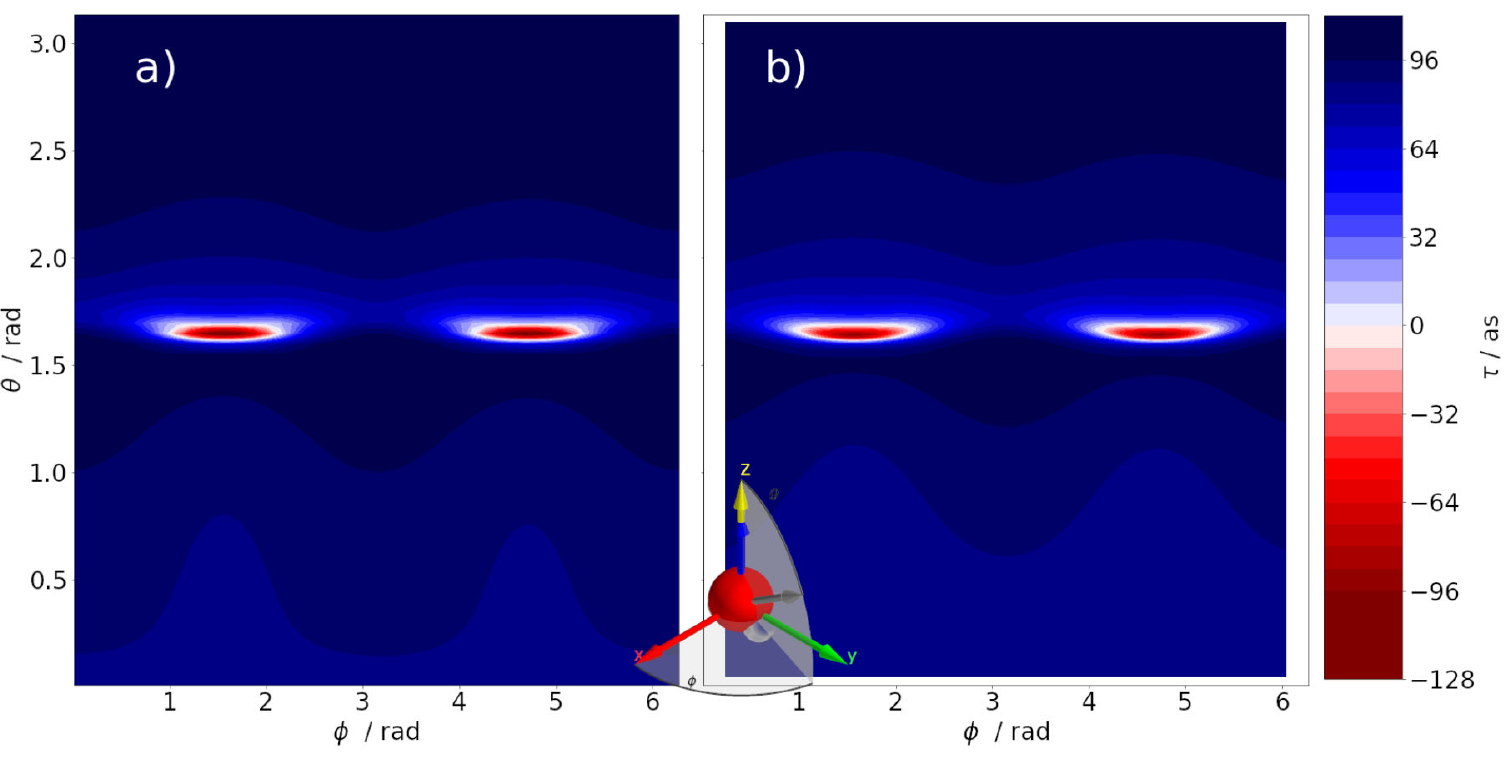}
\setcounter{figure}{9} 
\let\nobreakspace\relax
\renewcommand\figurename{\textbf{Fig. S}}
\def\fnum@figure{\figurename\thefigure}
\caption{\textbf{One-photon (Wigner) time delay a) and two-photon (XUV+IR) time delay b)} in the molecular frame as a function of electron-ejection angles ($\theta$ and $\phi$) in the molecular frame, when the light polarisation (XUV and IR, blue arrow) is aligned along the z-axis, as shown in the inset. All results are time delays calculated for the H$_2$O monomer, HOMO orbital and photoionization to SB12.}
\label{R10}
\end{figure}

\begin{figure}[h]
\centering
\includegraphics[width=0.7\textwidth]{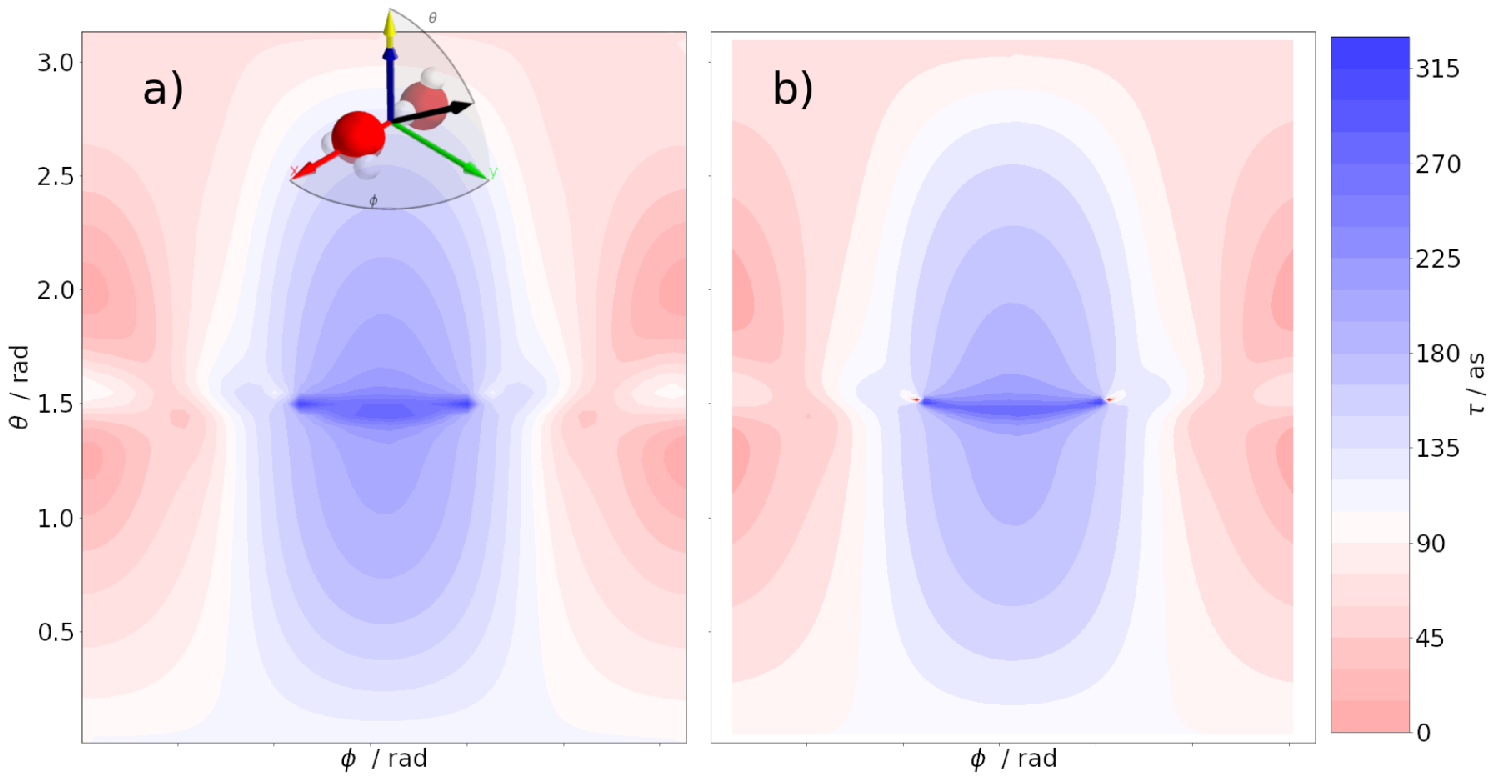}
\setcounter{figure}{10} 
\let\nobreakspace\relax
\renewcommand\figurename{\textbf{Fig. S}}
\def\fnum@figure{\figurename\thefigure}
\caption{\textbf{Same as Fig. S\ref{R10}} but for (H$_2$O)$_2$ and HOMO orbital.}
\label{R11}
\end{figure}

\begin{figure}[h]
\centering
\includegraphics[width=0.7\textwidth]{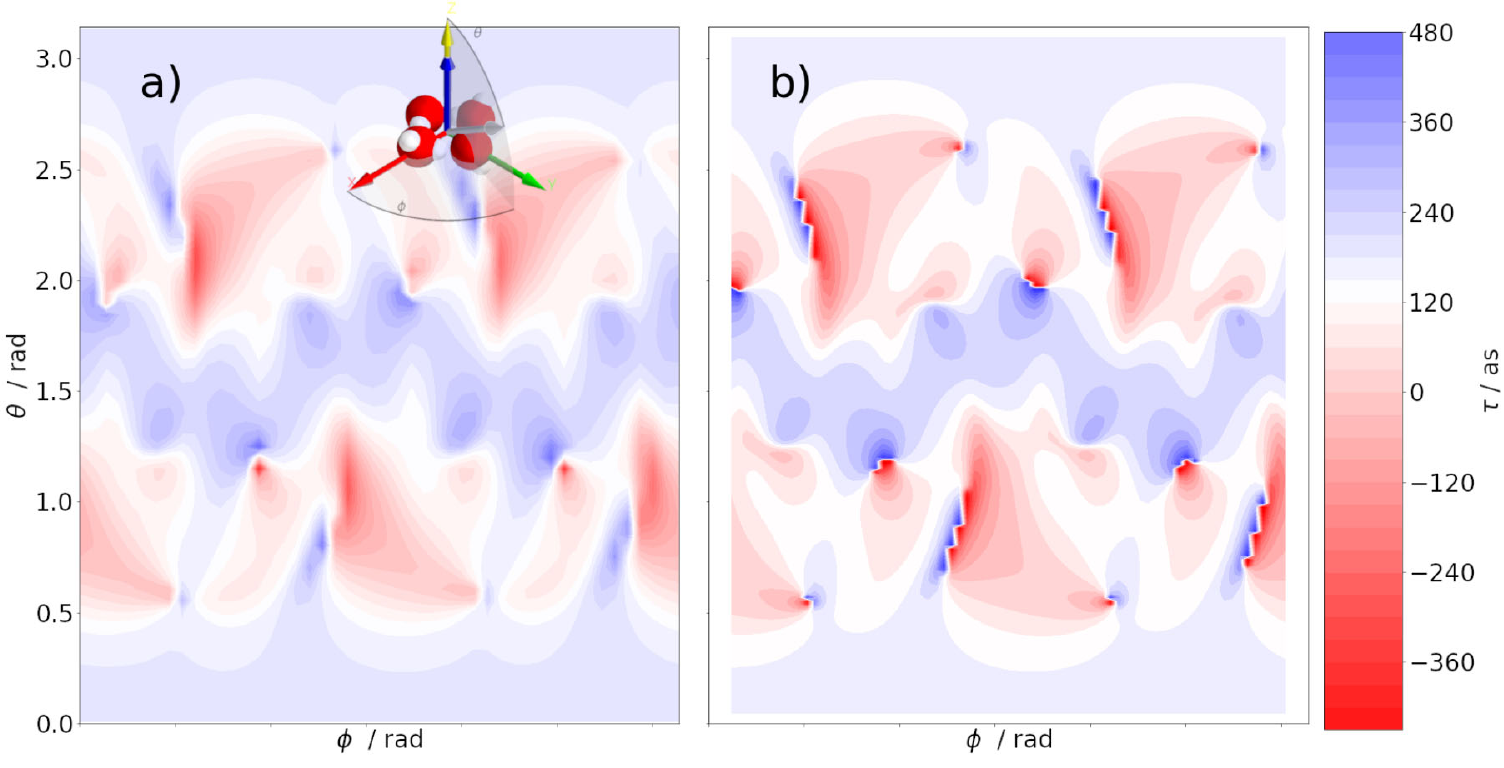}
\setcounter{figure}{11} 
\let\nobreakspace\relax
\renewcommand\figurename{\textbf{Fig. S}}
\def\fnum@figure{\figurename\thefigure}
\caption{\textbf{Same as Fig. S\ref{R10}} but for (H$_2$O)$_4^{(s4)}$ and HOMO orbital.}
\label{R12}
\end{figure}

Although the three-dimensional momentum is accessible in our coincidence setup, we averaged over all molecular orientations and the emission angle $\mathbf{\hat{\kappa}}$ of the photoelectron in the off-line analysis due to the limited water-cluster signal. Hence, the ionization time delay $\tau_{\omega}(E)$ is obtained by integrating over all molecular orientations (defined by the set of Euler angles $\hat R_\gamma$) and photoelectron-emission directions $\mathbf{\hat{\kappa}}$, as described in Ref.\cite{baykusheva17a}.
\begin{equation*}
  \tau_{\omega}(E)= \frac{1}{8\pi^2} \int d\hat R_\gamma  \int d\mathbf{\hat{\kappa}}  \,  A(E,\hat R_\gamma,\mathbf{\hat{\kappa}}) \,  \tau_{\omega}^{MF}(E,\hat R_\gamma,\mathbf{\hat{\kappa}})
  \label{eq.01}
\end{equation*}
where
\begin{equation*}
 A(E,\hat R_\gamma,\mathbf{\hat{\kappa}}) = \frac{|\sum_{lm\mu} I_{lm\mu} Y_{lm}(\mathbf{\hat{\kappa}}) D^{(1)}_{\mu m_p} (\hat R_\gamma)|^2 }{ \sum_{lm\mu} |I_{lm\mu}|^2 },
\end{equation*}
 both quantities being shown in Fig. S\ref{R9}.


Instead of directly measuring the one-photon (Wigner) delays $\tau_{\omega}(E)$, a RABBIT experiment accesses a closely related quantity, resulting from two-photon (XUV+IR) transitions. The molecular part of the two-photon time delay (ignoring cc-delay) in the molecular frame is defined via the total transition amplitude $f_{mol}^{MF}$ given by
\begin{equation*}
f_{mol}^{MF}(2q,\mathbf{\hat{\kappa}},\hat R_{\gamma}) =   \sum_{LM \\
L'M' } Y^*_{L'M'}(\mathbf{\hat{\kappa}}) Y_{LM}(\mathbf{\hat{\kappa}}) \, b^*_{L'M';(2q-1)}(\hat R_{\gamma}) \, b_{LM;(2q+1)}(\hat R_{\gamma}),
\end{equation*}
where transition amplitudes from IR emission/absorption $b$ have been defined in Ref.\cite{baykusheva17a}.
We get the laboratory-frame $\tau_{mol}(2q)$ time delay by coherently integrating $f_{mol}^{MF}$ over all molecular orientations (defined by the set of Euler angles $\hat R_\gamma$) and photoelectron-emission directions $\mathbf{\hat{\kappa}}$ and calculating the phase of the total transition amplitude, i.e.
\begin{equation*}
\tau_{mol}(2q) = \frac{1}{2\omega} \arg \left [\int d\hat R_\gamma \, \int d\mathbf{\hat{\kappa}}  \, f_{mol}^{MF}(2q,\mathbf{\hat{\kappa}},\hat R_{\gamma}) \right].
\end{equation*}
We can also calculate the molecular frame time delay from $f_{mol}^{MF}(2q,\mathbf{\hat{\kappa}},\hat R_{\gamma})$ directly,  and then proceed as in the one-photon case by cross-section-weighted integration (shown in RHS of Fig. S\ref{R10}, Fig. S\ref{R11} and Fig. S\ref{R12}). We have verified that the two approaches give the same results.

Figures S\ref{R10} to S\ref{R12} compare the molecular-frame one- and two-photon delays of the water clusters. These figures show that the two quantities are very similar in magnitude and display very similar angle dependencies. This agreement is further quantified in Fig. S\ref{R13}, which directly compares the one- and two-photon delays obtained after full angular averaging over the photoemission and molecular-orientation directions. In most cases, the difference between the delays amount to a few attoseconds only. The largest difference is found in the case of the S$_4$ isomer of the water tetramer, where the difference amounts to $\sim$13~as. Compared to the error bars of the experimental results, even this difference is still negligible.

\begin{figure}[h]
\centering
\includegraphics[width=0.7\textwidth]{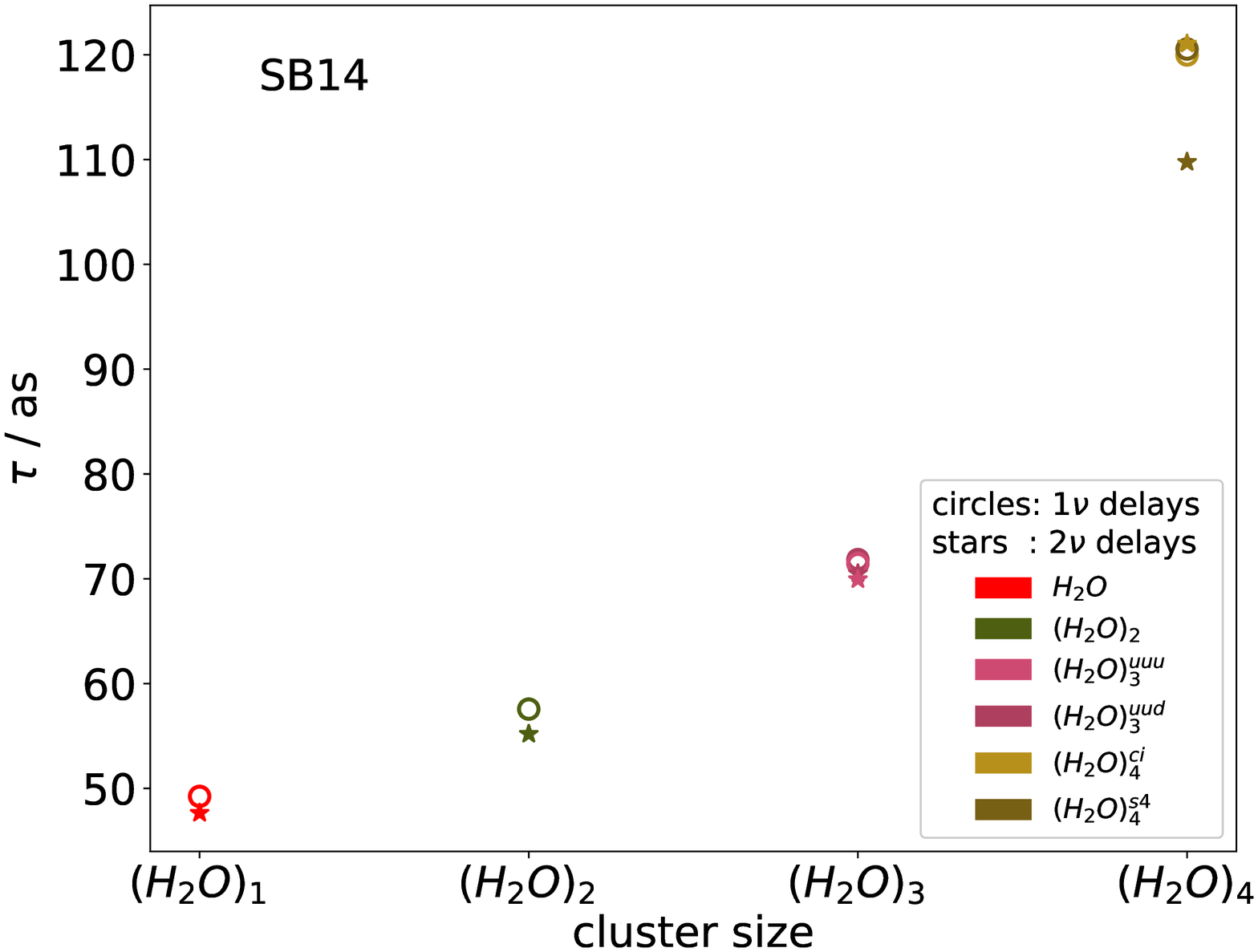}
\setcounter{figure}{12} 
\let\nobreakspace\relax
\renewcommand\figurename{\textbf{Fig. S}}
\def\fnum@figure{\figurename\thefigure}
\caption{\textbf{Comparison of absolute angle- and orientation-averaged one- and two-photon delays}  of water clusters in SB14. These results can be directly compared with the (relative) one-photon delays shown in Fig. 3b of the main text.
}
\label{R13}
\end{figure}

Based on this close agreement between one- and two-photon delays in water clusters, we have therefore based our discussion in the main text and in the remainder of this document on the one-photon (Wigner) delays.

\subsection*{2.3 Photoionization delays of the 1b$_1$ and O-1s (1a$_1$) bands of water clusters}

The following set of figures show the complete cluster-size-, isomer- and orbital-resolved one-photon-ionization delays integrated over target orientations and photoemission directions. A subset of these results are shown in Figs. 3 and 4 of the main text. Figure 3 of the main text shows the cross-section-weighted average time delay, which is defined as follows
\begin{equation}
    \tau(E)=\frac{\sum_{i=1}^n \sigma_i(E)\tau_i(E)}{\sum_{i=1}^n \sigma_i(E)},
\tag{S9}
\end{equation}
where the sum runs over the $n$ orbitals of the 1b$_1$ band of (H$_2$O)$_n$, $\sigma_i(E)$ is the photoionization cross section of orbital $i$ at the photon energy $E$ and $\tau_i(E)$ is the corresponding photoionization time delay. This quantity can directly be compared with the experimental results, which do not resolve the individual orbitals of the 1b$_1$ band.

Figure S14 shows the time delays and 1b$_1$-band orbital densities of the 1-2 most stable water-cluster isomers. A selection of these results is also shown in Figs. 3 and 4 of the main text. This figure highlights the close relationship between orbital delocalization and time delays. The largest delays are obtained for the tetramer with S$_4$ symmetry. This is consistent with the perfect orbital delocalization imposed by symmetry. This figure also nicely illustrates the generality of the localization phenomenon for clusters larger than the tetramer: all orbital densities for pentamers to heptamers are typically delocalized over $\sim$3 molecules. The delocalization does not augment with increasing cluster size.

Figure S15 shows the same results as Fig. S14, but displays the orbital wave functions, instead of the orbital densities. This representation highlights the additional possible role of the orbital sign. Comparing, as an example, HOMO and HOMO-3 of the tetramer-ci reveals that the "out-of-phase" combination of the 1b$_1$ molecular-fragment orbital has a much smaller delay (208~as) than the "in-phase" combination (420~as). The same trend can also be seen when comparing the HOMO (230~as) and HOMO-2 (425~as) of the tetramer-s4.

Figures S16 and S17 show the same orbital densities and wave functions as Figs.~S14 and S15, but provide the delays for a kinetic energy of 9.1~eV, corresponding to a photon energy of $\sim$21.7~eV, i.e. SB14.

Figure S18 shows the time delays and O-1s (or 1a$_1$) orbital densities of the same water-cluster isomers as Figs. S14-S17. A subset of these results is also shown in Figs. 3 and 4 of the main text. These calculations illustrate the direct relationship between time delay and orbital delocalization even more directly by removing the orbital hybridization effects caused by hydrogen bonding and orbital overlap in the 1b$_1$ band. The orbitals of the tetramer-ci are localized on a single water molecule and display a delay that is identical to the monomer. In contrast, the orbitals of the tetramer-s4 are perfectly delocalized (by symmetry) and all display much larger delays. In the O-1s band, the effect of disorder-induced orbital delocalization is even more striking. For all clusters larger than the tetramer, the orbitals are localized on individual molecules and the delays are close to the monomer delay. These results further confirm the negligible contribution of electron scattering off neutral water molecules to the time delays.

Figure S19 shows the same results as Fig. S18, but displays the orbital wave functions, instead of the orbital densities. In this case, the relationship between time delays and signs of the orbital wave functions appears to be opposite to the 1b$_1$ band. Comparing the HOMO and HOMO-3 of the tetramer-s4, it is the "out-of-phase" combination of the atomic O-1s orbitals that leads to the larger delay (141~as), compared to the "in-phase" combination. However, the less symmetric phase combination of HOMO-2 is associated with the largest delay (185~as). Therefore, we do not find a simple relationship between the signs of the linear combinations of fragment orbitals in clusters and the associated photoionization time delays.

\newpage
\begin{figure}
\includegraphics[width=0.9\textwidth]{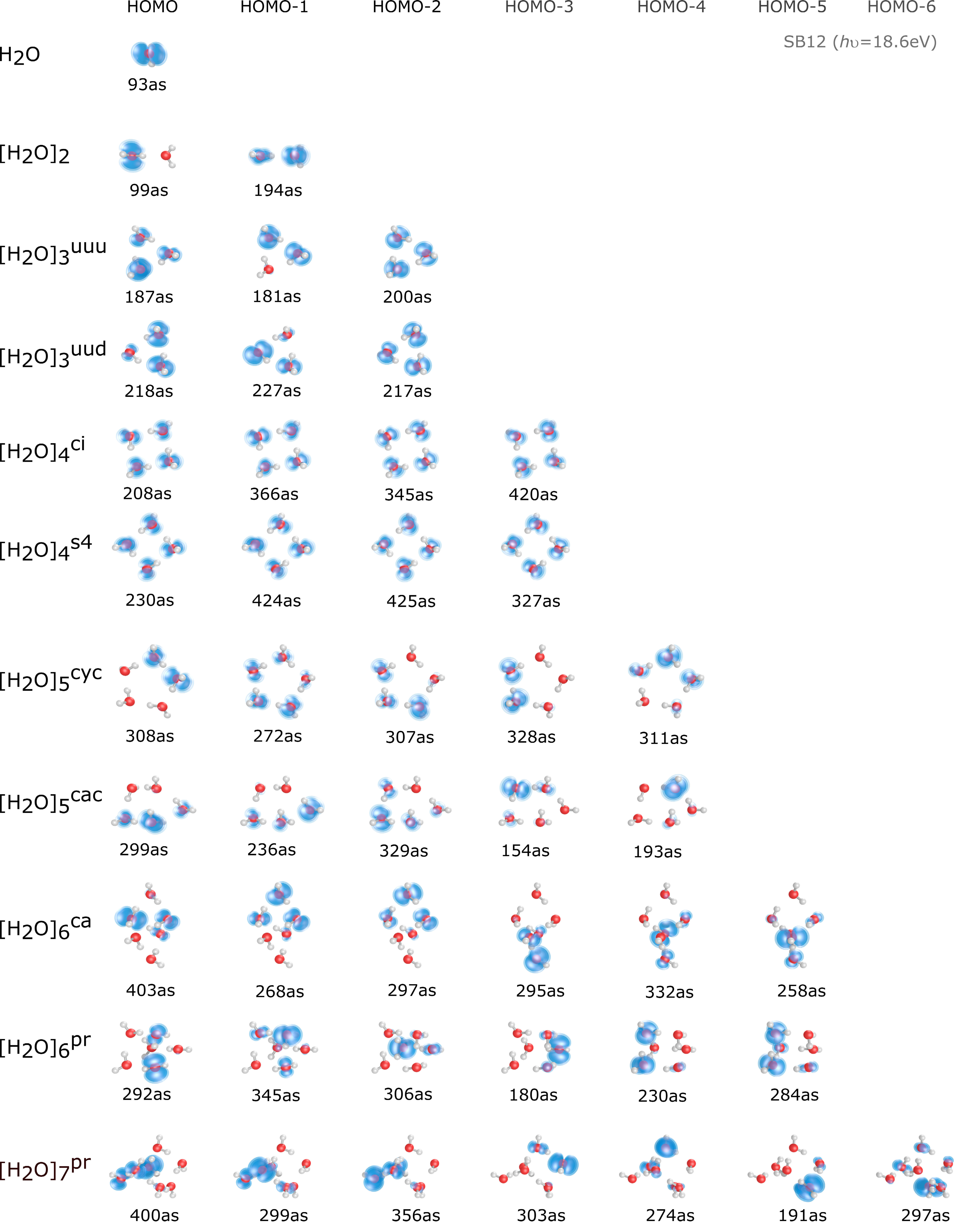}
\newline
{\textbf{Fig. S14: Orbital-resolved photoionization time delays of SB12 for the 1b$_1$ band.} Orbital density and photoionization time delays of water clusters using a kinetic energy of 6.0~eV, corresponding to a photon energy of $\sim$18.6~eV.}
\label{den_sb12}
\end{figure}

\newpage
\begin{figure}
\includegraphics[width=0.9\textwidth]{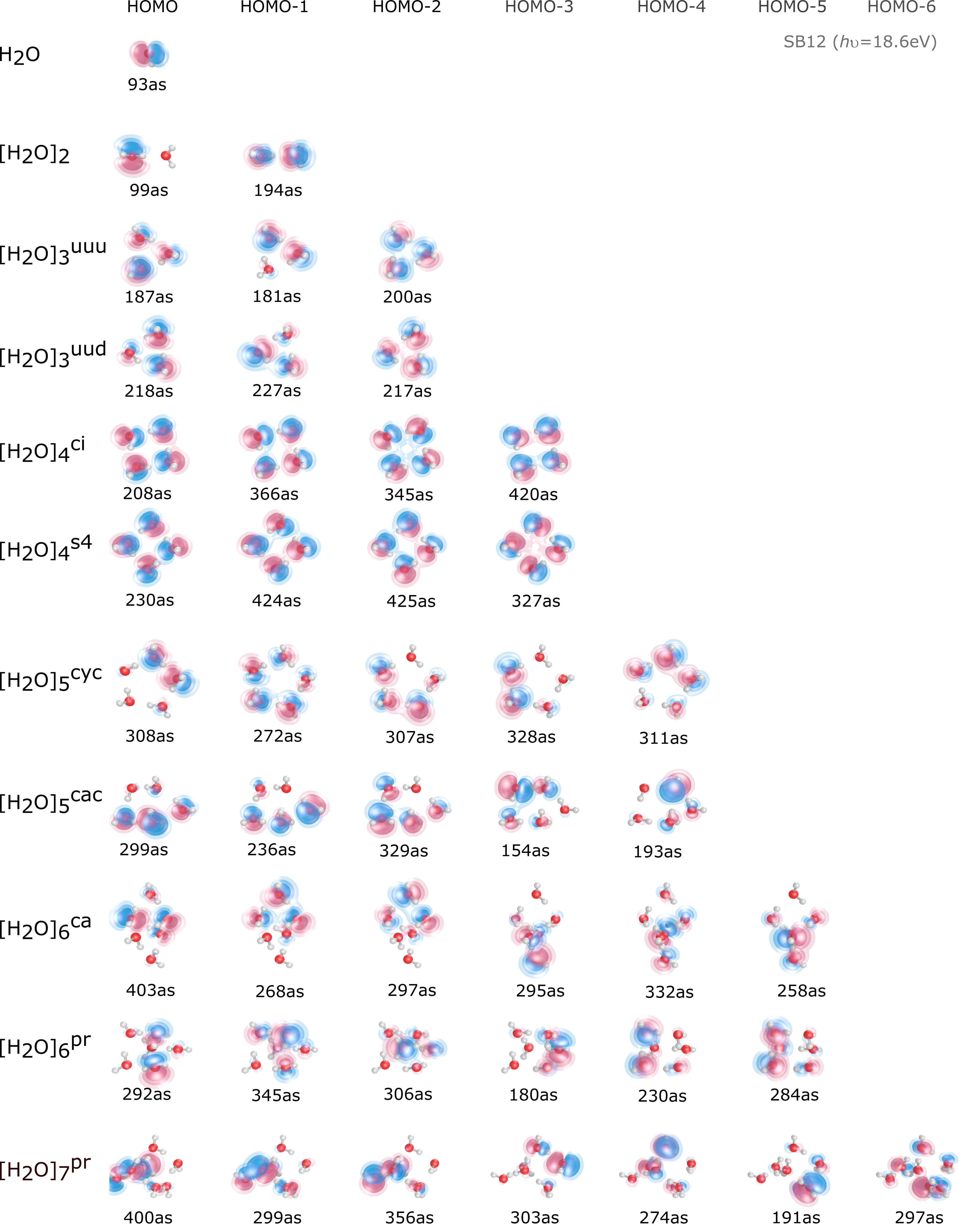}
\newline
{\textbf{Fig. S15: Orbital-resolved photoionization time delays of SB12 for the 1b$_1$ band.} Same as Fig. S14, but showing the orbital wave functions instead of their density.}
\label{wav_sb12}
\end{figure}

\newpage
\begin{figure}
\includegraphics[width=0.9\textwidth]{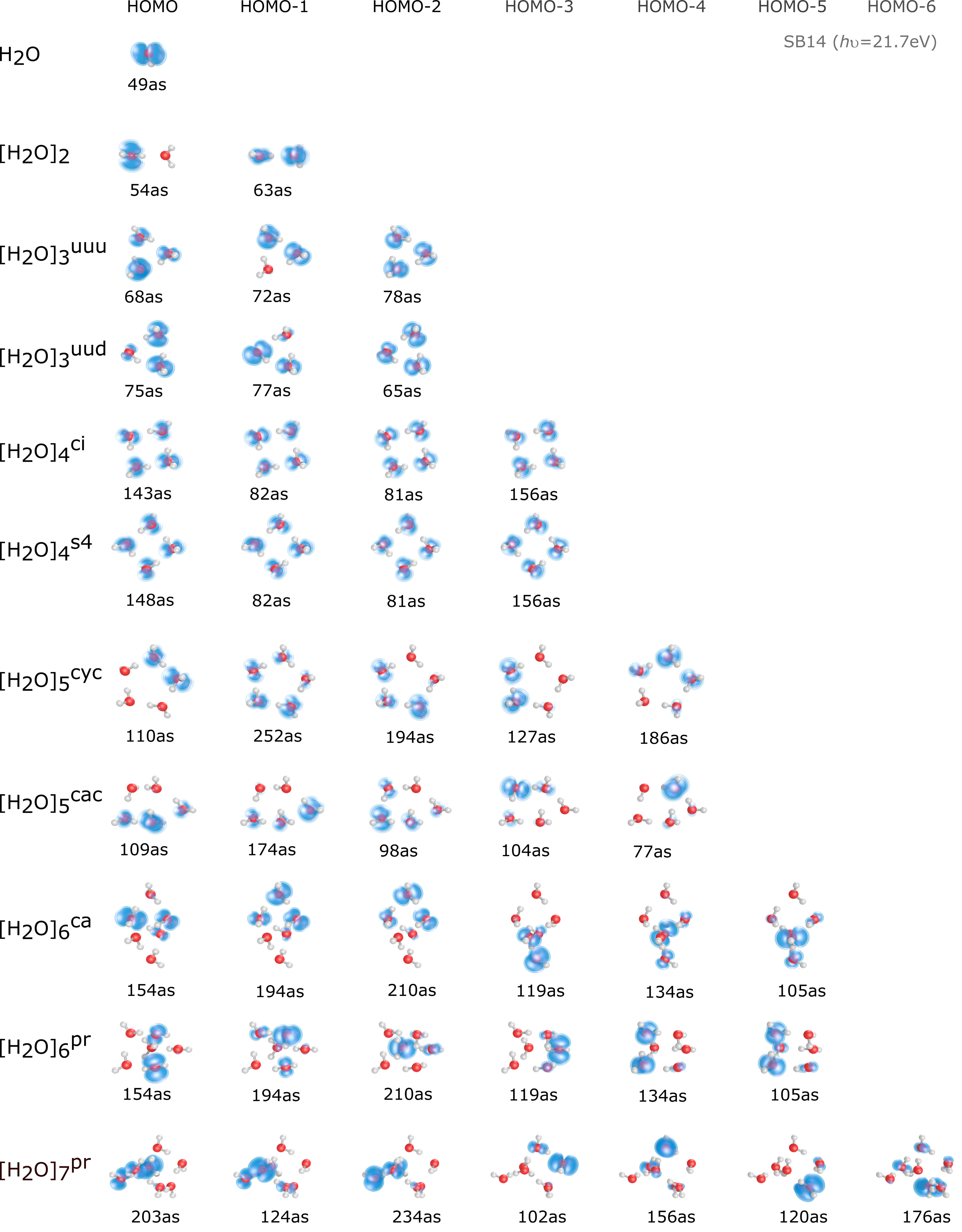}
\newline
{\textbf{Fig. S16: Orbital-resolved photoionization time delays of SB14 for the 1b$_1$ band.} Orbital density and photoionization time delays of water clusters using a kinetic energy of 9.1~eV, corresponding to a photon energy of $\sim$21.7~eV.}
\label{den_sb14}
\end{figure}

\newpage
\begin{figure}
\includegraphics[width=0.9\textwidth]{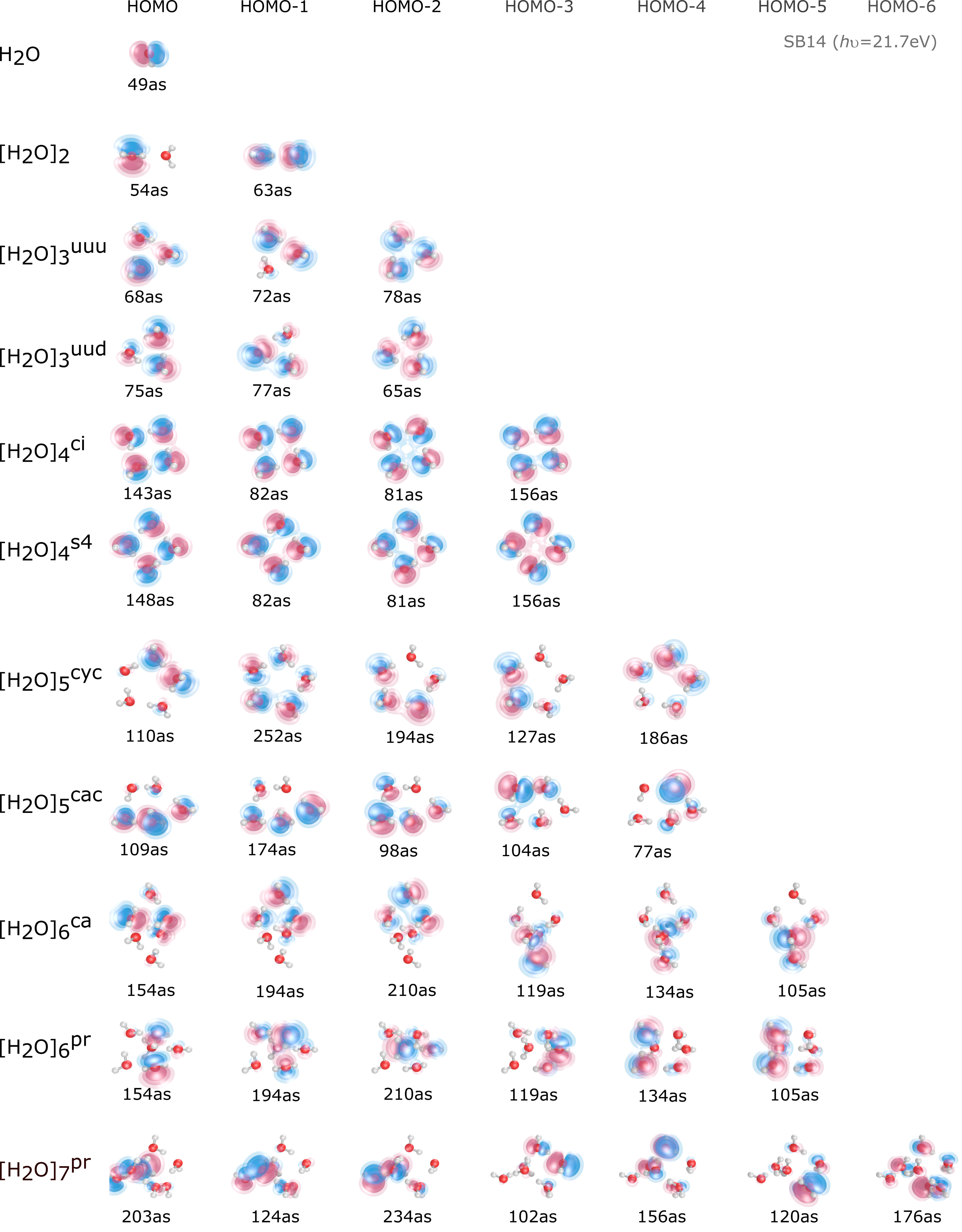}
\newline
{\textbf{Fig. S17: Orbital-resolved photoionization time delay of SB14 for the 1b$_1$ band.}  Same as Fig. S16, but showing the orbital wave functions instead of their density.}
\label{wav_sb14}
\end{figure}

\newpage
\begin{figure}
\includegraphics[width=0.9\textwidth]{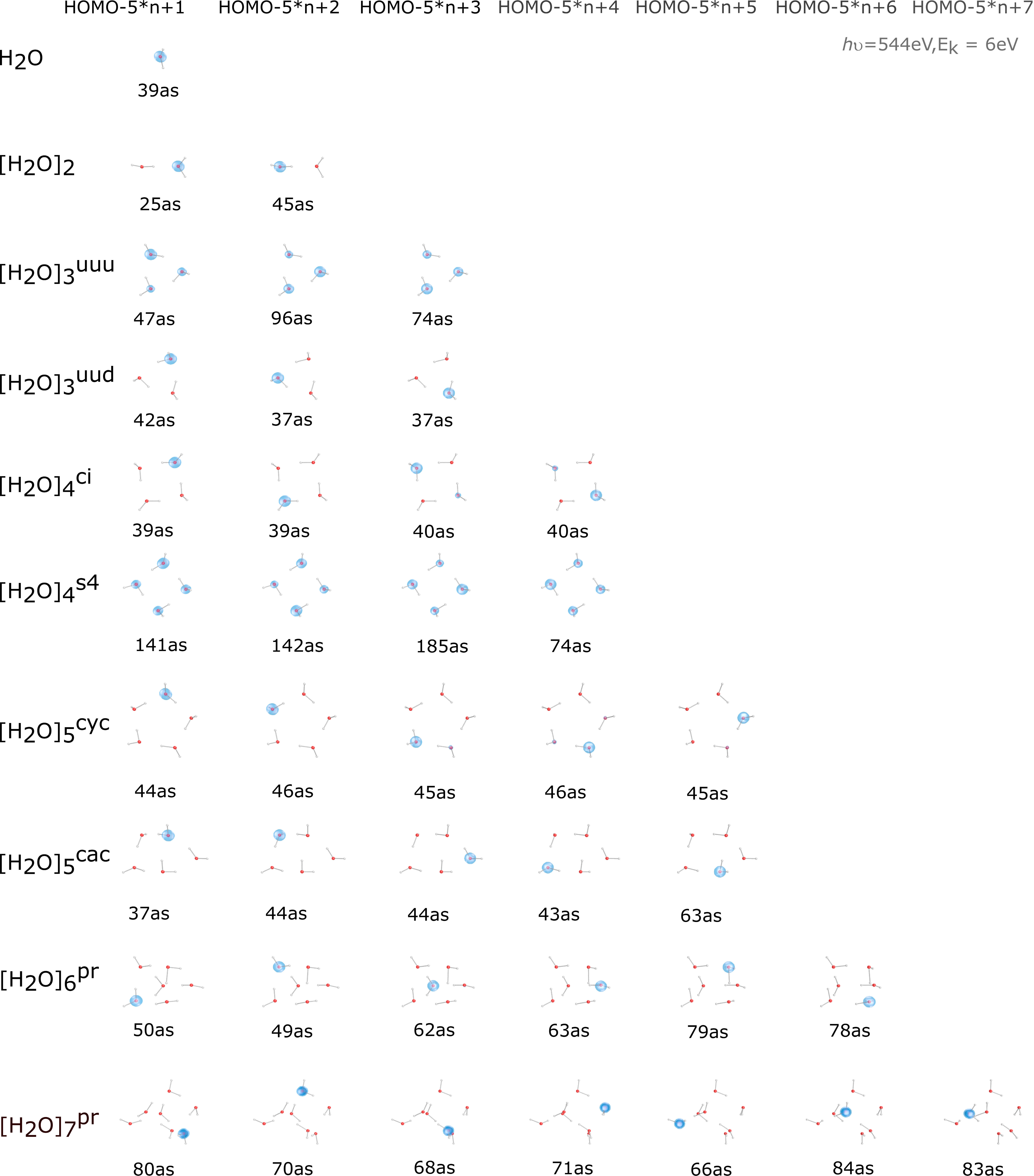}
\newline
{\textbf{Fig. S18: Orbital-resolved photoionization time delays of the O1s (or 1a$_1$) band.} Orbital density and photoionization time delays of water clusters using a kinetic energy of 6.0~eV, corresponding to a photon energy of $\sim$544~eV.}
\label{den_o1s}
\end{figure}

\newpage
\begin{figure}
\includegraphics[width=0.9\textwidth]{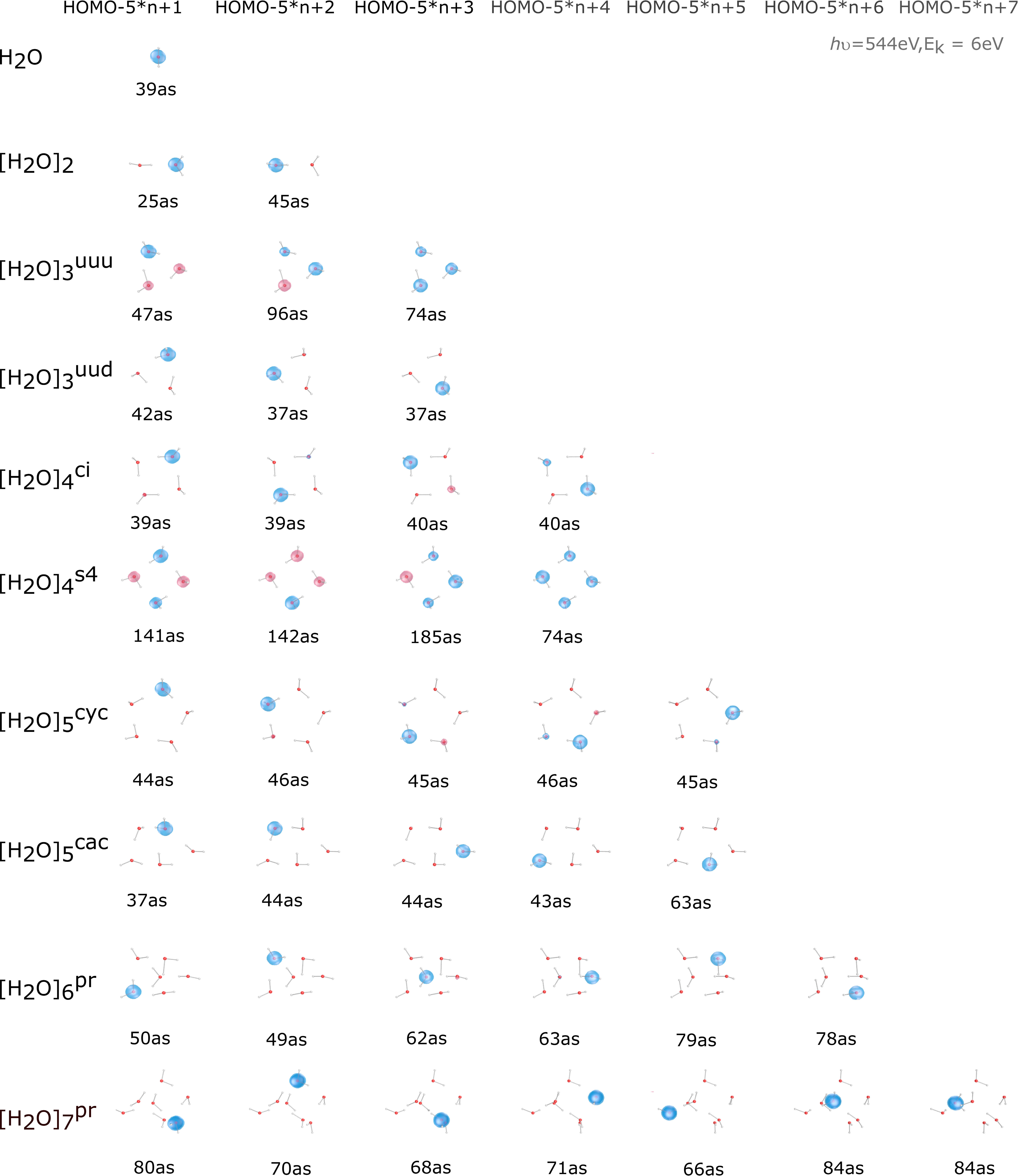}
\newline
{\textbf{Fig. S19: Orbital-resolved photoionization time delays of the O1s (or 1a$_1$) band.} Same as Fig. S18, but showing the orbital wave functions instead of their density.}
\label{wav_o1s}
\end{figure}

\clearpage
\newpage

\subsection*{2.4 Effects of nuclear motion and resonances on photoionization delays of water clusters}

In this section, we briefly discuss the possible role of nuclear motion on photoionization delays of water clusters. Ionization of a water cluster initiates the transfer of a proton from the ionized molecule to its neighbor, as illustrated in Fig. S20. The time scale for proton transfer in ionized water clusters has recently been determined to be $\sim$30-40 fs \cite{svoboda20a}.

\begin{figure}
\begin{center}
\includegraphics[width=0.7\textwidth]{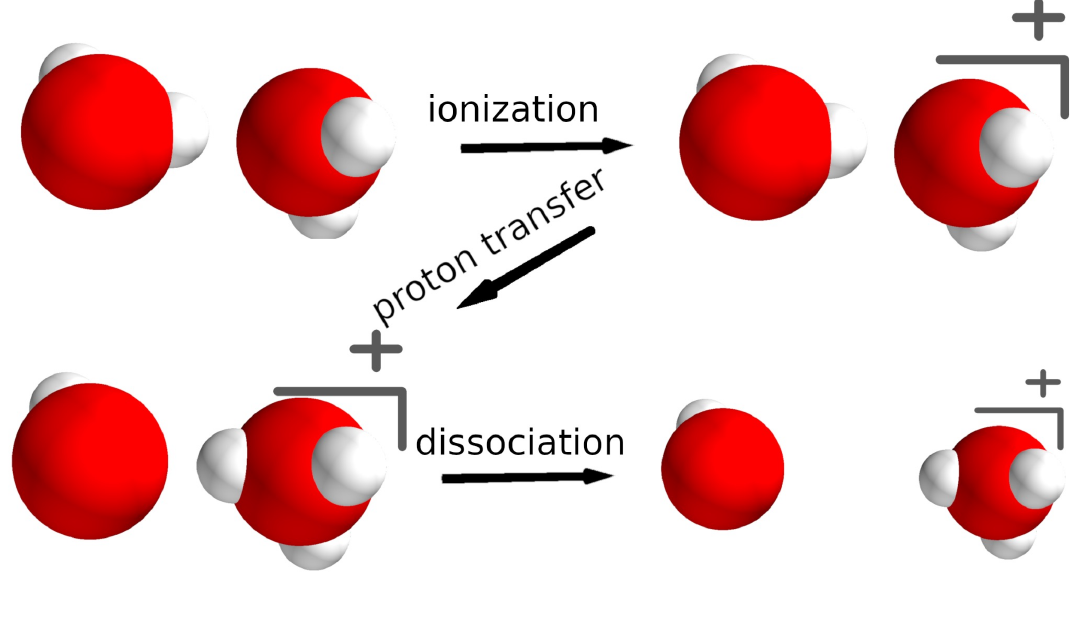}
\end{center}
{\textbf{Fig. S20: Illustration of the large amplitude motion (proton transfer) following ionization of water clusters, illustrated for the example of the water dimer.} A similar illustration for the case of the water trimer is shown in Fig. 1b of the main text.}
\end{figure}

To assess the role of nuclear motion on the photoionization time delays, we have calculated the variation of the time delays along the proton-transfer coordinate, as well as the O-O stretch coordinate in the water dimer. Figure S21 shows that the photoionization delays, obtained as a function of these nuclear coordinates vary very little (by $\pm$3~as) across the Franck-Condon regions, shown as red-shaded areas. These results support the fixed-nuclei approximation used in the remainder of our work.

\begin{figure}
\begin{center}
\includegraphics[width=1.0\textwidth]{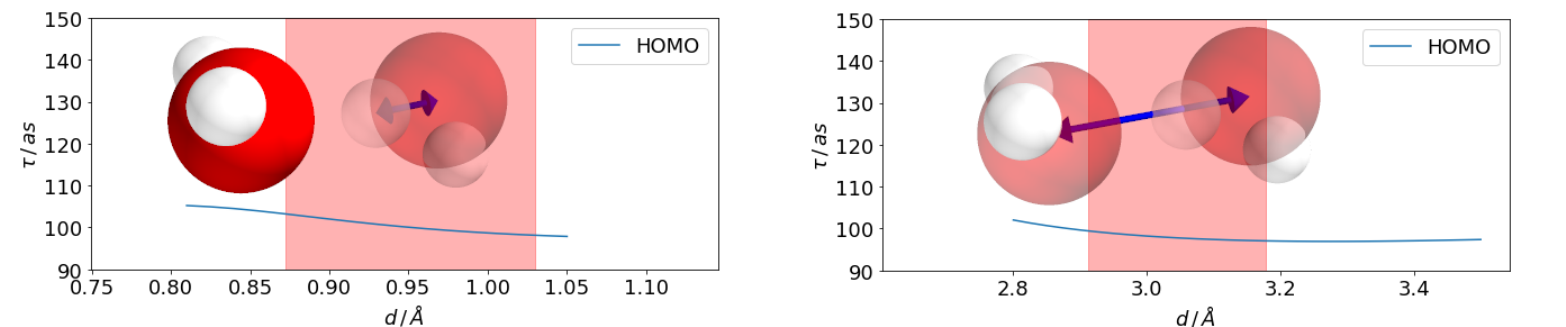}
\end{center}
{\textbf{Fig. S21: Photoionization time delays for the HOMO of the water dimer as a function of OH bond length (left) and OO bond length (right) for the case of SB12.} The shaded area is the Franck-Condon (FC) region for the vibrational ground state of (H$_2$O)$_2$.}
\end{figure}

We moreover studied the energy dependence of the photoionization time delays with the goal of locating possible shape resonances or Cooper(-like) minima. Previous work has shown that shape resonances usually lead to pronounced local maxima in the photoionization delays \cite{huppert16a,baykusheva17a,nandi20a}, whereas Cooper minima give rise to local minima \cite{Schoun2014}. In the case of water clusters, we find that the variation of the photoionization time delays is strictly monotonic in the energy region of interest in our work, which indicates the absence of shape resonances or Cooper-like minima.

\begin{figure}
\begin{center}
\includegraphics[width=0.48\textwidth]{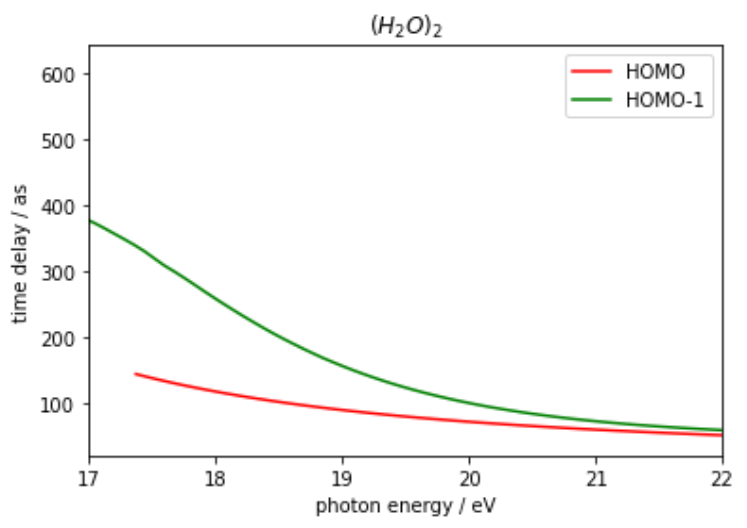}
\includegraphics[width=0.48\textwidth]{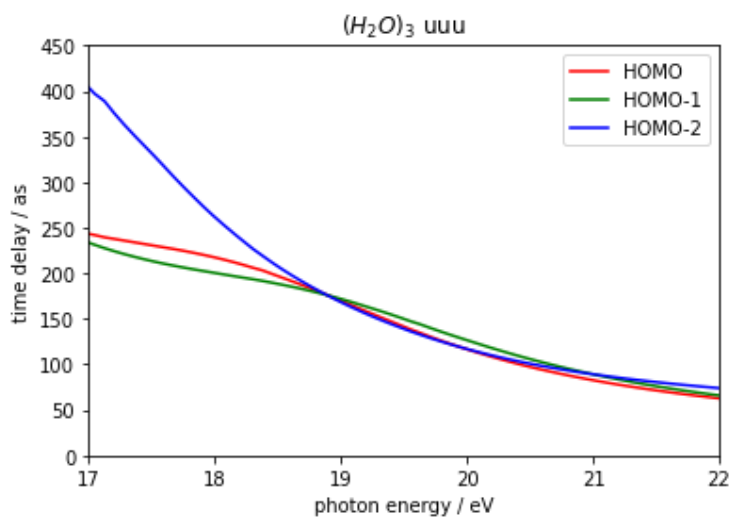}
\end{center}
{\textbf{Fig. S22: Orbital-resolved photoionization time delays of individual orbitals of the 1b$_1$ band of the water dimer (left) and trimer (right).} }
\label{wav_o1s_2}
\end{figure}

\subsection*{2.5 Benchmarking the quantum-scattering calculations against experimental photoelectron asymmetry parameters}
Finally, we present an independent validation of our quantum-scattering calculations by comparison with an independent experimental observable, i.e. the photoelectron asymmetry parameters of water clusters\cite{Uwe2013,hartweg2017a}. Figures S23 compares the measured photoelectron asymmetry parameters to our calculations at the kinetic energies corresponding to SB12 and SB14. The asymmetry parameters were obtained from the output of ePolyScat \cite{Lucchese94,Lucchese99}, which uses Eq. (12) in \cite{Lucchese99} for the calculation of the asymmetry parameters of individual orbitals. We subsequently performed a cross-section-weighted average of the $\beta$ values of the individual orbitals $i=1-N$ of the 1b$_1$-band of (H$_2$O)$_n$ according to
\begin{equation}
    \beta(E)=\frac{\sum_{i=1}^n \sigma_i(E)\beta_i(E)}{\sum_{i=1}^n \sigma_i(E)},
\end{equation}
The very good agreement between measured asymmetry parameters and those calculated at the electron-kinetic energies corresponding to SB12 and SB14 in our work independently validates the accuracy of our quantum-scattering calculations on water clusters.

\begin{figure}
\begin{center}
\includegraphics[width=0.8\textwidth]{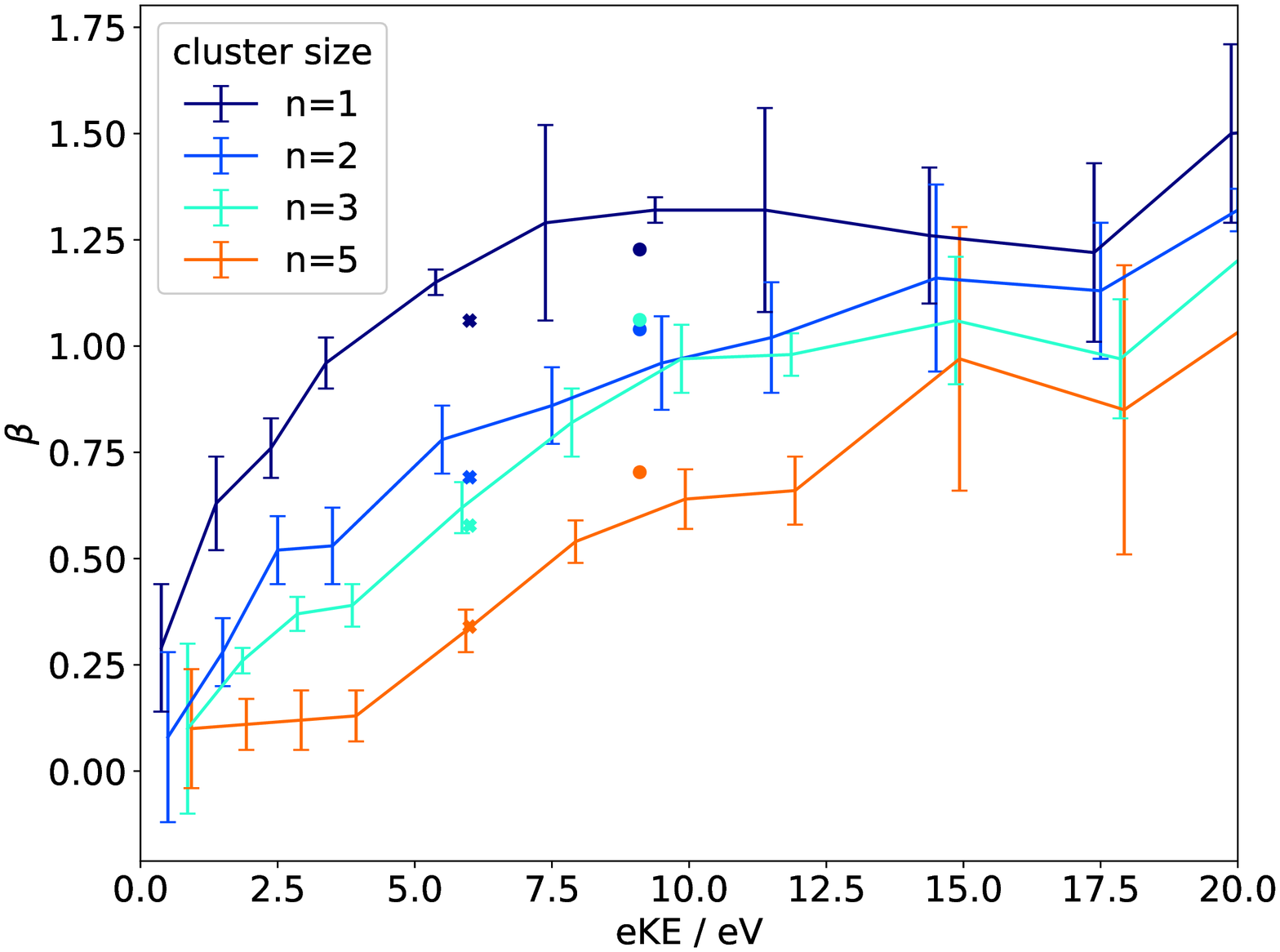}
\end{center}
{\textbf{Fig. S23: Asymmetry parameters ($\beta$) for photoionization from the 1b$_1$-band of water clusters.} Comparison of experimental (lines, Ref.\cite{hartweg2017a}) and calculated (crosses/circles, this work) asymmetry parameters ($\beta$) as a function of the electron kinetic energies.}
\end{figure}



\clearpage
\newpage
\bibliography{attoH2On,attobib}%
\bibliographystyle{naturemag_noURL}